\documentclass[conference]{IEEEtran}
\IEEEoverridecommandlockouts
\usepackage{fancyhdr}
\usepackage{cite}
\usepackage{amsmath,amssymb,amsfonts}

\usepackage{booktabs}
\usepackage{makecell}
\usepackage[utf8]{inputenc}
\usepackage{colortbl}
\usepackage{amsmath,amsfonts}
\usepackage{algorithm}
\usepackage{algorithmicx}
\usepackage{multirow}
\usepackage[noend]{algpseudocode}
\usepackage{algorithm}
\usepackage{graphicx}
\usepackage{textcomp}
\usepackage{url}
\PassOptionsToPackage{hyphens}{url}\usepackage[hidelinks]{hyperref}
\usepackage[dvipsnames]{xcolor}
\usepackage{forest, tikz}
\usetikzlibrary{decorations.pathreplacing,calligraphy}
\usepackage{enumitem}
\usepackage{subcaption}
\usepackage{sidecap}
\usepackage{sc26repro}
\usepackage{fvextra}
\usepackage{fancyvrb}

\usepackage{pgfplots}
\pgfplotsset{compat=1.18}

\usepackage{tikz}
\usetikzlibrary{positioning,fit,calc,arrows.meta}
\usepackage{xcolor}

\usepackage{tikzscale}
\usetikzlibrary{positioning,shapes.geometric,arrows.meta}
\usepackage{listings}
\usepackage{bm}

\usepackage{comment}

\usepackage{tikz}
\usepackage{xcolor}
\usepackage{xparse}
\usetikzlibrary{calc}
\usetikzlibrary{matrix}
\usetikzlibrary{arrows.meta}
\usetikzlibrary{backgrounds}
\usepackage{xparse}

\usepackage{xspace}    
\usepackage{tabularx}

\usepackage[disable]{todonotes} % disable for testing page length

\newcommand{\revanth}[1]{\todo[color=yellow, inline]{Revanth: #1}}

\newcolumntype{Y}{>{\centering\arraybackslash}X}
\newcolumntype{R}{>{\raggedleft\arraybackslash}X}
\newcolumntype{L}{>{\raggedright\arraybackslash}X}

\def\susy{\textit{SuSy}\xspace} 
\def\higgs{\textit{Higgs}\xspace} 
\def\bigcross{\textit{BigCross}\xspace} 
\def\wave{\textit{WEC}\xspace} 
\def\msd{\textit{MSD}\xspace} 
\def\sift{\textit{Sift}\xspace} 

\def\iono{\textit{Ionosphere}\xspace} 
\def\expo{\textit{Exponential}\xspace} 
\def\uniform{\textit{Uniform}\xspace} 
\def\gaia{\textit{Gaia}\xspace}

\def\dss{DSS\xspace} 

\def\ouralg{\textsc{RT-HiSS}\xspace} 
\def\mikealg{\textsc{GDS-Join}\xspace}
\def\brianalg{\textsc{COSS}\xspace}
\def\cuvsalg{\textsc{cuVS-Brute}\xspace}
\def\GTS{\textsc{GTS}\xspace}
\def\pyTorchAlg{\textsc{PyTorch3D}\xspace}

\def\rtnn{\textsc{RTNN}\xspace}

\def\optix{\textsc{OptiX}\xspace}
\def\shq{\texttt{Shared-\allowbreak Query}\xspace}
\def\shp{\texttt{Shared-\allowbreak Primitive}\xspace}
\def\shno{\texttt{Shared-\allowbreak None}\xspace}
\def\shqp{\texttt{Shared-\allowbreak Query-\allowbreak Primitive}\xspace}

\def\timeout{\textsc{TimeOut}\xspace}

\usepackage{pifont}% http://ctan.org/pkg/pifont
\newcommand{\cmark}{\ding{51}}%
\newcommand{\xmark}{\ding{55}}%

\begin{document}

\title{RT-HiSS: Ray Tracing Accelerated High Dimensional Vector Similarity Searches \thanks{This material is based upon work supported by the National Science Foundation under Grant No. 2042155.}}

\author{\IEEEauthorblockN{Revanth Reddy Munugala}
\IEEEauthorblockA{\textit{School of Informatics, Computing, and Cyber Systems} \\
\textit{Northern Arizona University}\\
Flagstaff, AZ, U.S.A. \\
rm2878@nau.edu}
\and
\IEEEauthorblockN{Michael Gowanlock}
\IEEEauthorblockA{\textit{School of Informatics, Computing, and Cyber Systems} \\
\textit{Northern Arizona University}\\
Flagstaff, AZ, U.S.A. \\
michael.gowanlock@nau.edu}
}

\maketitle

% \thispagestyle{fancy}
% \lhead{}
% \rhead{}
% \chead{}
% \lfoot{\footnotesize{
% SC26, November 15-20, 2026, Chicago, Illinois, USA
% \newline 979-8-3195-4789-7/26/\$31.00 \copyright 2026 IEEE}}
% \rfoot{}
% \cfoot{}
% \renewcommand{\headrulewidth}{0pt}

\begin{abstract}

Recent GPU generations include special-purpose ray tracing (RT) cores for graphics applications. While RT cores are primarily used for rendering, recent works show they can be leveraged for general-purpose tasks, including similarity searches. However, existing approaches do not support datasets exceeding three dimensions. In this work, we propose \ouralg, the first exact GPU RT-core-based similarity search algorithm for high-dimensional datasets. GPU similarity search often scales poorly for large datasets with substantial search distances. To address this, \ouralg uses RT cores for fast index construction and searches, followed by candidate refinement on CUDA cores. We introduce a two-pass approach to estimate an upper bound on result size, enabling efficient batching under GPU memory constraints with near-perfect load balancing. Additionally, we examine shared memory tiling and compressed result masks to improve GPU resource utilization. \ouralg yields speedups up to 8.37$\times$ over competitive state-of-the-art GPU algorithms across six real-world datasets.

% compared to five state-of-the-art GPU algorithms.

% across six real-world datasets.

\end{abstract}


\section{Introduction}\label{Sec:intro}

% \mike{Mike todo: discuss HPC systems and discuss kernel-level optimizations}

Current scientific instruments generate enormous quantities of data that need to be processed, and consequently there has been significant effort to efficiently compute data analysis workflows on parallel and high performance computing  (HPC) systems. In this paper, we examine distance similarity searches (\dss) which are a data analytic method that are used for numerous tasks, including clustering~\cite{schubert2017dbscan}, outlier detection~\cite{mandhare2017comparative}, and natural language processing~\cite{mathur2024vector}, among others. \dss on points refer to identifying points in a dataset that are similar to a given query point based on a similarity metric. These searches are computationally expensive as they require numerous distance comparisons between query and data points.

HPC systems are becoming increasingly heterogeneous and nodes often contain two or more processor types, including  CPUs, Graphics Processing Units (GPUs), and Data Processing Units (DPUs). Even GPUs themselves are equipped with both tensor and ray tracing cores~\cite{nvidia2025blackwell}. An example of a highly heterogeneous HPC system is Frontera which contains both Nvidia V100 and RTX 5000 GPUs to address FP32 and FP64 workloads~\cite{stanzione2020frontera}, and of particular interest to this paper is that the RTX 5000 GPUs contain ray tracing (RT) cores. 

Several papers have examined kernel-level optimizations for \dss on GPUs~\cite{gdsJoin, coss, zhu2022rtnn, GTS, donnelly2024multi} including those using GPU RT cores because they benefit from hardware-accelerated searches on Bounding Volume Hierarchy (BVH) trees~\cite{zhu2022rtnn,nagarajan2023rt}. These similarity search algorithms employ two phases: \emph{$(i)$ An index search phase} that prunes the search to identify candidate neighbors that may be within the search distance ($\epsilon$) of query points. \noindent\emph{$(ii)$ A refinement phase} that compares query and candidate neighbor points to identify the exact neighbors within $\epsilon$. The search-and-refine approach using RT cores yields two major challenges, as follows.

% here is an opportunity to exploit them for \dss because they benefit from hardware-accelerated indexes.  

% Furthermore, instead of using GPU ray tracing (RT) cores for their primary purpose (rendering graphics in computer games and simulations) 

% RT cores are generally used to render graphics in computer games and simulations, among other applications~\cite{zhu2022rtnn,nagarajan2023rt}. 

\noindent\textbf{Challenge \#1: Unconstrained Result Set Sizes: }
% \mike{I think we need to shorten the two challenges. I'm working on this.}
\dss identify \emph{all} neighbors within $\epsilon$ of a given query point. The number of neighbors identified for each query is a function of dataset properties (i.e., distribution, number of points, and dimensionality). Subsequently, the result set size often exceeds GPU global memory capacity when processing large datasets and typically batching is employed to address this limitation. We identify several unintended consequences of batching which include: $(i)$ poor saturation of GPU resources due to insufficient work in each batch; $(ii)$ load imbalance across batches; and $(iii)$ overhead costs~\cite{gowanlock2019accelerating,donnelly2024multi}.

% \revanth{Should we say RT similarity search algorithms represent the search space of each data point as a primitive. I know you do not like saying search space. So, should we say ``local neighborhood" or ``$\epsilon$-neighborhood. Applies to all material throughout the paper.}

\noindent\textbf{Challenge \#2: Limitations of RT Cores: }
RT cores are limited to datasets having  $\le 3$ dimensions; therefore, they cannot be directly employed for high-dimensional datasets~\cite{zhu2022rtnn, evangelou2021fast}. RT similarity search algorithms typically represent each data point as an individual geometrical primitive and use it to construct a BVH tree to efficiently prune the data space~\cite{zhu2022rtnn, nagarajan2023rt}. 
However, this approach results in a large number of ray-primitive intersections during BVH tree traversals, particularly for large datasets which significantly deteriorates performance. 

% RT similarity search algorithms exclusively use RT cores for refinement, and do not necessarily exploit the best GPU processor type (e.g., using CUDA cores) for each phase~\cite{zhu2022rtnn, evangelou2021fast}. 

% \revanth{Should we add the word ``scalable" in the first contribution?}

We demonstrate that using RT cores for indexing on three dimensions in conjunction with CUDA cores for refinement addresses the challenges above. To our knowledge, we propose the first algorithm to compute \dss on high-dimensional datasets using RT cores. In summary, this paper makes the following contributions:

\begin{itemize}[noitemsep,topsep=0pt,parsep=0pt,partopsep=0pt,leftmargin=*,wide]
    \item We propose a scalable RT accelerated high dimensional vector similarity search algorithm (\ouralg) by indexing on RT cores, and refining candidate neighbors using CUDA cores.
    \item We identify nearby data points, and represent the $\epsilon$-neighborhood of all such data points as a single primitive in the \optix scene to mitigate ray-primitive intersections. 
    \item We design a multi-pass ray-tracing approach that yields the upper bound on the number of neighbors for each query and a batching scheme to mitigate global memory limitations.
    \item We propose a method for compressing the result set by representing neighbors on the bit level. This method increases GPU utilization across batches.
    % \item We examine differing methods of representing the result set and compresses the result set size by representing neighbors on the bit level. This method increases GPU utilization across batches.
    % \item We employ a result mask with a lossless compression strategy to facilitate fast data transfers from global memory (GPU) to main memory (CPU). 
    % \item We propose a shared memory tiling approach to mitigate slow global memory accesses when refining candidate neighbor points.
\end{itemize}

Paper organization: Section~\ref{sec:background} outlines background material. Section~\ref{sec:optixModel} details the \optix programming model. Section~\ref{sec:ourAlg} describes our algorithm, \ouralg. Section~\ref{sec:expEval} presents the evaluation. Section~\ref{sec:conclusion} concludes the paper.

\section{Background}\label{sec:background}

% We present the problem statement and related work.

We present the problem statement and discuss the limitations of RT cores for distance similarity searches in Section~\ref{sec:problemStatement}. We detail the prior work on indexing methods used in \dss, as well as GPU-based approaches in Section~\ref{sec:relatedWork}.

\subsection{Problem Statement}\label{sec:problemStatement}
Instead of treating \dss as a classical problem that relies on spatial partitioning methods, we reformulate it as a ray tracing problem.\footnote{To avoid confusion, \dss in the literature are also referred to as fixed radius searches, $\epsilon$-neighborhood searches, or proximity searches. We use the term \dss throughout the paper.} Although prior work has used this approach~\cite{zhu2022rtnn, evangelou2021fast}, these works are limited to datasets with $\le 3$ dimensions (Section~\ref{sec:relatedWork}). In this work, we propose Ray Tracing Accelerated High Dimensional Vector Similarity Search (\ouralg) algorithm for large high dimensional datasets.

In a dataset $D$ having $d$ dimensions, we define the coordinates (or vector values) of a point $p_i\in D$ as $p_i=\{x_i^1, x_i^2, \ldots, x_i^d\}$, where $i=1, 2, \ldots, |D|$. Similarly, we define query set $Q$, having query points $q_j \in Q$ where $j = 1, 2, \ldots,|Q|$.

% \revanth{Should we say something like ``We employ Euclidean distance throughput the paper as it is widely used in other studies~\cite{xxx} xx."}

The Euclidean distance between two points $p_i \in D$ and $q_j \in Q$ is defined as $dist(p_i, q_j) = \sqrt{\sum_{k=1}^{d} ({x_i}^k - {x_j}^k)^2}$. Two points $p_i, q_j$ are defined as \emph{neighbors} of each other if and only if $dist(p_i, q_j) \leq \epsilon$, where $\epsilon$ is the search distance. We employ the Euclidean distance throughout the paper as it is widely used in other studies~\cite{wang2025euclidean, Astronomyreview}.

In database terminology, the case above is referred to as a \emph{semi-join}, where $D$ and $Q$ are distinct. If $Q=D$ then this is referred to as a self-join where the data points themselves are also the query points, which is common in many data analytic applications (inlier/outlier detection, clustering, among others~\cite{orair2010distance,kulkarni2024survey}). We evaluate our algorithm using the \emph{self-join} scenario, but the algorithm and optimizations are directly applicable to the semi-join as well.

%,gowanlock2018hybrid,munugala2024gdbod

% In this work, we primarily investigate the \emph{self-join} scenario, where the query set is the same as the dataset ($Q=D$). Nevertheless, our proposed algorithm and all the optimizations in this work are applicable to semi-join as well.

\subsubsection{Result Set Representation}\label{sec:resultRep}

As described above, the points that are within $dist(p_i, p_j)\leq\epsilon$ are stored in the result set, that we denote as $R$. However, the representation of $R$ varies throughout the literature. The most conservative approach is to store a distance matrix of size $|Q|\times|D|$~\cite{li2015brute} such that for each query point, $q_j\in Q$, there is sufficient storage to indicate that all points in the dataset have been found by a given query. A more practical approach is to simply store key-value pairs, where the key is the query point ID for $q_j\in Q$ and the value is the point ID of $p_i\in D$~\cite{gdsJoin}. This stores the total number of neighbors that are within the $\epsilon$ of each other, where there are a total of $|R|$ results and thus key-value storage requires $2\cdot|R|$ elements. Comparing the distance matrix to key-value pairs, the latter requires significantly less storage for real-world application scenarios. Thus, in our evaluation, we compare our approach to storing the result set to the key-value pair approach instead of the distance matrix.

\subsubsection{RT Cores} 

We represent the $\epsilon$-neighborhood  of data points in the dataset $D$ as geometrical primitives in an \optix scene in our algorithm, where the \optix scene represents the data space. RT cores are limited to 3 dimensions because only 3 spatial dimensions are needed for ray tracing. To enable searches in high dimensions, we use RT cores to index on three dimensions ($d=3$) to identify candidate neighbor points for each query and later refine these candidate points using conventional CUDA cores in a separate kernel. 

% A detailed explanation of \ouralg is presented in Section~\ref{sec:ourAlg}.

\subsection{Related Work}\label{sec:relatedWork}

The literature on distance similarity searches is extensive. Searching each query point using a linear scan over the dataset is computationally expensive and yields an $O(|D|^2)$ time complexity and is referred to as a brute-force approach. The performance of \dss can be improved by: $(i)$ Employing an index to partition the search space and limit the number of distance calculations between query and data points.  $(ii)$ Leveraging specialized hardware, particularly GPUs to accelerate distance calculations as they are well-suited for data parallel tasks like \dss due to their Single Instruction Multiple Data (SIMD) architecture. We discuss these two approaches below.

% In what follows, we describe indexing methods that prune the search space to reduce the number of distance comparisons and thus mitigate the quadratic time complexity.

% \mike{Expand this a bit, put tree-based on a new line and introduce the indexing methods in the sunbsubsection}
\subsubsection{Indexing Methods}

Indexing methods prune the search space and reduce the number of distance comparisons and thus mitigate the quadratic time complexity. In what follows, we describe various indexing methods used in the literature.

\noindent\textbf{Tree-based:} Tree-based data structures such as kd-trees~\cite{kdTree}, R-trees~\cite{guttman1984r}, Oct-trees~\cite{jackins1980oct}, and Ball-trees~\cite{dolatshah2015ball} prune the search space through hierarchical partitioning, enabling efficient searches. However, tree-based algorithms suffer from poor performance on high-dimensional datasets due to the \emph{curse of dimensionality}~\cite{bellman1961adaptive}, where the search space increases exponentially with data dimensionality, and so a corresponding increase in search distance, $\epsilon$, is needed to find nearby neighbors. In tree-based approaches, this often implies that a large fraction of the tree needs to be searched which often degrades to a brute-force search where all pairs of points are compared to each other.

\noindent\textbf{Grid-based:} Grid-based algorithms employ coordinate-based indexing to partition the data space using axis-aligned partitions to optimize the search by pruning the data space and considering at most $3^d$ non-empty neighboring grid cells~\cite{gdsJoin}. 

\noindent\textbf{Metric-based:} Metric-based algorithms use reference points in the data space and leverage the distance between each data point and its closest reference point to prune the search space. This strategy is typically more effective than grid-based indexing in high-dimensional spaces, but incurs higher indexing costs in low-dimensional spaces~\cite{donnelly2024multi}.

All of the above indexing methods have their strengths and weaknesses and their performance largely depends on the dataset properties such as the dimensionality, distribution, and size, $|D|$. As such, in our evaluation, we compare our approach to several indexes having varying properties, including grids, trees, coordinate- and metric-based approaches.

\subsubsection{GPU Similarity Searches}
We briefly review several GPU similarity search algorithms in the literature. As described in the  papers below, they are well-suited for GPU architectures as queries can be executed independently in massively parallel fashion.

\noindent\textbf{GPU CUDA Core–based \dss:}
\textsc{PCLOctree} is an implementation from the Point Cloud Library (PCL) that supports kNN and \dss~\cite{PCL}. Grid-based indexing approaches include \mikealg~\cite{gdsJoin} and \textsc{FRNN}~\cite{FRNN}. Metric-based approaches that use reference points (or pivot points) include \GTS~\cite{GTS}, \brianalg~\cite{coss}, and \textsc{MiSTIC}~\cite{donnelly2024multi}.
All of these approaches rely only on general-purpose CUDA cores for pruning the search space and identifying neighbors, and do not leverage RT cores.

\noindent\textbf{GPU RT Core–based \dss:}
Evangelou et al.~\cite{evangelou2021fast} proposed the first similarity search algorithm using RT cores. Zhu proposed~\textsc{RTNN}~\cite{zhu2022rtnn}, which uses query partitioning and Z-order sorting to perform efficient neighbor searches using RT cores. Meneses et al.~\cite{menesesRT} presented a similarity search algorithm that determines whether the BVH tree should be updated or reconstructed from scratch when performing updates. Lastly, while Nagarajan et al.~\cite{nagarajan2023rt} did not conduct \dss, they proposed $k$-nearest neighbor searches using RT cores, which is an orthogonal algorithm but its methods overlap with \dss.

\emph{All existing RT core similarity search algorithms are limited to datasets with $d\leq 3$ due to the hardware constraints. We address this limitation and to our knowledge, we are the first to exploit RT cores for high dimensional \dss where $d\gg3$.}

\section{\optix Programming model}\label{sec:optixModel}

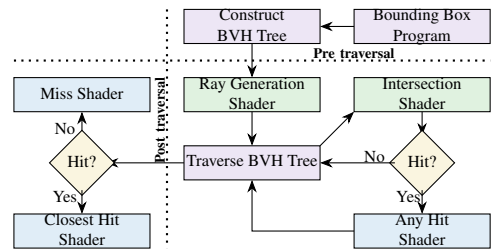
\begin{figure}[!t]
\centering
        \def\scale{0.45}
\definecolor{lightblue}{rgb}{0.88,0.93,0.97}
\definecolor{lightgreen}{rgb}{0.88,0.95,0.88}
\definecolor{lightyellow}{rgb}{0.98,0.96,0.88}
\definecolor{lightpurple}{rgb}{0.93,0.90,0.96}

\begin{tikzpicture}[scale=\scale, transform shape]

\draw[thin, fill=lightpurple] (0,2) rectangle (4,3);
\node at (2,2.5) {\Large{\makecell{Construct \\ BVH Tree}}};
\draw[arrows={-Stealth}] (2,2) -- (2,1);

\draw[thin, fill=lightpurple] (5,2) rectangle (9,3);
\node at (7,2.5) {\Large{\makecell{Bounding Box \\ Program}}};
\draw[arrows={-Stealth}] (5,2.5) -- (4,2.5);

\draw[thin, fill=lightgreen] (0,0) rectangle (4,1);
\node at (2,0.5) {\Large{\makecell{Ray Generation \\ Shader}}};
\draw[arrows={-Stealth}] (2,0) -- (2,-1);

\draw[thin, fill=lightpurple] (0,-2) rectangle (4,-1);
\node at (2,-1.5) {\makecell{ \Large{Traverse BVH Tree}}};
\draw[arrows={-Stealth}] (4,-1) -- (5,0);
\draw[arrows={-Stealth}] (0,-1.5) -- (-2.3,-1.5);

\draw[thin, fill=lightgreen] (5,0) rectangle (9,1);
\node at (7,0.5) {\Large{\makecell{Intersection\\ Shader }}};
\draw[arrows={-Stealth}] (7,0) -- (7,-0.7);

\node[draw, diamond, aspect=1, fill=lightyellow] at (7,-1.5) {\Large{\makecell{Hit?}}};
\draw[arrows={-Stealth}] (6.2,-1.5) -- (4,-1.5);
\draw[arrows={-Stealth}] (7,-2.3) -- (7,-3);

% \draw[thin] (5,-4) rectangle (9,-3);
\draw[thin, fill=lightblue] (5,-4) rectangle (9,-3);
\node at (7,-3.5) {\Large{\makecell{Any Hit \\ Shader}}};
\draw[arrows={-Stealth}] (5,-3.5) -- (2,-3.5) -- (2,-2);

\draw[thick, dotted] (-0.5,3) -- (-0.5,-4);
\draw[thick, dotted] (-5,1.5) -- (9,1.5);

\node at (5,1.7) {\large{\textbf{Pre traversal}}};

\node[rotate=90] at (-0.7,-0.25) {\large{\textbf{Post traversal}}};

\draw[arrows={-Stealth}] (-3,-0.7) -- (-3,0);
\node[draw, diamond, aspect=1, fill=lightyellow] at (-3,-1.5) { \Large{\makecell{Hit?}}};
\draw[arrows={-Stealth}] (-3,-2.3) -- (-3,-3);

\node at (6.6, -2.5) {\Large{Yes}};
\node at (5.6, -1.3) {\Large{No}};

\draw[thin, fill=lightblue] (-5,0) rectangle (-1,1);
\node at (-3,0.5) {\Large{\makecell{Miss Shader}}};

\node at (-3.5, -2.5) {\Large{Yes}};
\node at (-3.5, -0.5) {\Large{No}};

\draw[thin, fill=lightblue] (-5,-4) rectangle (-1,-3);
\node at (-3,-3.5) {\Large{\makecell{Closest Hit \\ Shader}}};

\end{tikzpicture}
        \caption{\optix programming model~\cite{NvidiaDocs}. Shaders used in \ouralg are shown in green, while the rest are in blue.}
        \label{fig:RTModel}
\end{figure}

NVIDIA introduced the \optix API in 2009, enabling users to harness ray tracing capabilities with support for user-defined geometries using programmable shaders that execute during different stages of the ray tracing pipeline~\cite{NvidiaDocs, zhu2022rtnn}. Figure~\ref{fig:RTModel} illustrates the \optix programming model~\cite{NvidiaDocs}.

BVH tree is constructed using the bounding volumes of user-defined geometries. The ray tracing pipeline starts with ray generation where Ray Generation Shader launches rays based on the ray origin and direction. During the BVH tree traversal, the Intersection Shader is triggered whenever a ray intersects with an object in the scene. After an intersection is detected, the Any-Hit Shader enables users to define custom processing based on intersection data. Similarly, the Closest-Hit Shader processes the closest intersection, while the Miss Shader handles cases where a ray does not intersect any objects in the scene~\cite{NvidiaDocs}. We only use Ray Generation and Intersection shader in \ouralg.

\subsection{Distance Similarity Searches using RT Cores}\label{sec:rangeSearches}

\begin{figure}[!t] 
  \centering
  % \revanth{Need to update formatting for captions.}
  \setlength{\tabcolsep}{4pt}
  % \revanth{Fix figure}
  \begin{tabularx}{\columnwidth}{lL}
  
   % \documentclass[tikz,border=10pt]{standalone}
% \usepackage{tikz}
% \usetikzlibrary{positioning,fit,calc,arrows.meta}
% \usepackage{xcolor}

% \usepackage{tikzscale}
% % Check if there are duplicates
% \usetikzlibrary{positioning,shapes.geometric,arrows.meta}
% \begin{document}

\begin{tikzpicture}[scale=0.75]

    \coordinate (P3) at (-2.25,1.45);
    
    \fill (P3) circle (1pt);
    \draw[thin] (P3) circle (1);
    \node[above right] at (P3) {\tiny $p_3$};
    \draw[densely dashed, very thin] (-3.25,0.45) rectangle (-1.25,2.45);
    
    \node at (-1.55,2.55) {\tiny $AABB_3$};
    
    \fill (0,0) circle (1pt);
    \draw[thin] (0,0) circle (1);
    \node[below left] at (0,0) {\tiny $p_1$};
    \draw[densely dashed, very thin] (-1,-1) rectangle (1,1);
    \node at (-0.60,1.10) {\tiny $AABB_1$};

    \draw[-Stealth, very thin] (0,0) -- (0.7,-0.7);
    \node at (0.20,-0.35) {\tiny $\epsilon$};
       
    \fill (1,1) circle (1pt);
    \draw[thin] (1,1) circle (1);
    \node[above right] at (1,1) {\tiny $p_2$};
    \draw[densely dashed, very thin] (0,0) rectangle (2,2);
    \node at (1.65,-0.15) {\tiny $AABB_2$};

    \fill (0.50,0.50) circle (1pt);
    \node[below] at (0.5,0.5) {\tiny $p_q$};

    \node at (-3.15, 0.20) {\tiny \textcolor{red}{$BV_3$}};
    \draw[densely dotted, thin, red] (-3.35,0.35) rectangle (-1.15,2.55);

    \node at (-0.80, 1.95) {\tiny \textcolor{red}{$BV_{12}$}};
    \draw[densely dotted, thin, red] (-1.10,-1.10) rectangle (2.10,2.10);

     \draw[densely dotted, thin, blue] (-3.4,-1.15) rectangle (2.15,2.60);
     \node at (1.80, 2.45) {\tiny \textcolor{blue}{$BV_{123}$}};

\end{tikzpicture}

% \end{document} 
   &
    \begin{tikzpicture}[scale=0.80]

% ===== Layer 0 =====
\draw[densely dotted, thin, blue] (0,0) rectangle (1,0.5);
\node at (0.5,0.25) {\textcolor{blue}{\tiny $BV_{123}$}};

\draw[-Stealth, very thin] (0.5,0) -- (-0.6,-0.5);
\draw[-Stealth, very thin] (0.5,0) -- (1.6,-0.5);

% ===== Layer 1 =====
% BV_3
\draw[densely dotted, thin, red] (-1.1,-1) rectangle (-0.1,-0.5);
\node at (-0.6,-0.75) {\textcolor{red}{\tiny $BV_{3}$}};

% BV_12
\draw[densely dotted, thin, red] (1.1,-1) rectangle (2.1,-0.5);
\node at (1.6,-0.75) {\textcolor{red}{\tiny $BV_{12}$}};

\draw[-Stealth, very thin] (-0.6,-1) -- (-0.6,-1.5);
\draw[-Stealth, very thin] (1.6,-1) -- (1.1,-1.5);
\draw[-Stealth, very thin] (1.6,-1) -- (2.5,-1.5);

% ===== Layer 2 =====
% AABB_3
\draw[densely dashed, thin] (-1.1,-2) rectangle (-0.1,-1.5);
\node at (-0.6,-1.75) {\tiny $AABB_{3}$};

% AABB_1
\draw[densely dashed, thin] (0.8,-2) rectangle (1.8,-1.5);
\node at (1.3,-1.75) {\tiny $AABB_{1}$};

% AABB_2
\draw[densely dashed, thin] (2.0,-2) rectangle (3.0,-1.5);
\node at (2.5,-1.75) {\tiny $AABB_{2}$};

\draw[-Stealth, very thin] (-0.6,-2) -- (-0.6,-2.5);
\draw[-Stealth, very thin] (1.3,-2) -- (1.3,-2.5);
\draw[-Stealth, very thin] (2.5,-2) -- (2.5,-2.5);

% ===== Layer 3 =====
\draw (-0.6,-2.75) circle (0.25);
% \fill (-0.6,-3) circle (1pt);
\node at (-0.6,-2.75) {\tiny $p_3$};

\draw (1.3,-2.75) circle (0.25);
% \fill (1.3,-3) circle (1pt);
\node at (1.3,-2.75) {\tiny $p_1$};

\draw (2.5,-2.75) circle (0.25);
% \fill (2.5,-3) circle (1pt);
\node at (2.5,-2.75) {\tiny $p_2$};

\end{tikzpicture}
% \end{document}
 \\
    
   \footnotesize{ \textbf{\makecell[l]{ (a) Bounding volumes and AABBs \\ of $\epsilon$-neighborhoods of points $p_1$, \\ $p_2$, and $p_3$.}}}
   & 
  \footnotesize \textbf{\makecell[l]{(b) BVH tree constructed \\using bounding volumes \\for DSS.}} \\
 \end{tabularx}
 \caption{\dss using RT cores. Here, data points $p_1$ and $p_2$ are candidate neighbor points for query point $q_j$.}
 \label{fig:neighborSearch}
\end{figure}
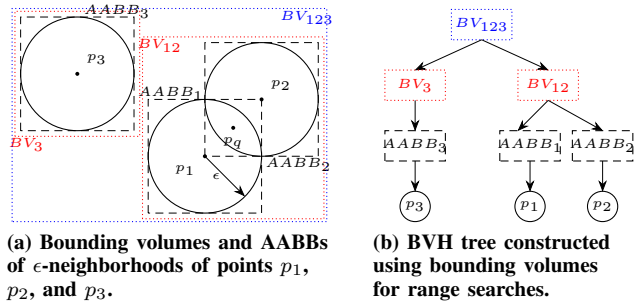

To identify all neighbors in a dataset within a search distance $\epsilon$  of a query point $q_j$, data points are represented in the scene using a user defined geometry which is often a sphere for $d=3$ searches. These user defined geometries or geometrical primitives are represented by Axes Aligned Bounding Boxes (AABBs). The AABBs tightly encompass the primitives and are used to construct a BVH tree. The \optix BVH tree construction details are abstracted from the programmer. Similar to the prior work~\cite{zhu2022rtnn, nagarajan2023rt} that employed RT cores for similarity and $k$-nearest neighbor searches, to identify all the neighbors of a query point $q_j$, we invoke an infinitesimally short ray in any arbitrary direction with the origin as the query point $q_j$. We use an infinitesimally short ray to avoid false positives, as a long ray could potentially intersect many distant AABBs in the scene. Instead of checking if each point $p_i \in D$ falls within the $\epsilon$-neighborhood  of $q_j$, we perform an inverse search. We check if the query point $q_j$ falls within the $\epsilon$-neighborhood of each data point in the dataset $D$.
% \mike{Nice explanation!}
% \revanth{Add one line explaining why the second method is more efficient.}

% \mike{where circles represent the search space changed to eps neighborhood}
Figure~\ref{fig:neighborSearch} shows how neighbors are identified using RT cores. Figure~\ref{fig:neighborSearch}(a) illustrates a two-dimensional space, where circles represent the $\epsilon$-neighborhood of data points $p_1$, $p_2$, and $p_3$, respectively. The black dotted lines indicate the AABBs encompassing the $\epsilon$-neighborhood of each point, which are used to construct the BVH tree. To effectively prune unnecessary ray–primitive interactions, the BVH tree groups nearby bounding volumes and recursively constructs higher level bounding volumes that enclose these groups. In Figure~\ref{fig:neighborSearch}(a), $BV_{12}$ represents the bounding volume enclosing $AABB_1$ and $AABB_2$, which are close to each other, while $BV_3$ represents the bounding volume enclosing $AABB_3$. Finally, $BV_{123}$ encloses all bounding volumes in the scene.

% \mike{Update the query point $q_j$ notation}
Figure~\ref{fig:neighborSearch}(b) illustrates the BVH tree constructed using the bounding volumes in Figure~\ref{fig:neighborSearch}(a). Here, $BV_{123}$ is the root node encompassing all bounding volumes, $BV_3$ and $BV_{12}$ are intermediate nodes, and the leaf nodes are $AABB_3$, $AABB_1$, and $AABB_2$. The leaf nodes encompass the $\epsilon$-neighborhood  of points $p_1$, $p_2$, and $p_3$, respectively. Hence, any ray invoked from the query point $q_j$ that intersects the corresponding AABB implies that the query point could potentially be within the $\epsilon$-neighborhood  of the corresponding data point.

% \mike{Update the query point $q_j$ notation, in the caption and here, and potentially elsewhere.}
To identify the neighbors of the query point $q_j$, we traverse the BVH tree shown in Figure~\ref{fig:neighborSearch}(b). The invoked infinitesimally short ray intersects the bounding volume $BV_{12}$. Therefore, the AABBs within $BV_{12}$ (i.e., $AABB_1$ and $AABB_2$) are considered as candidate neighbors for $q_j$. The bounding volume $BV_3$ is ignored since the ray does not intersect it. The same principle applies to large numbers of data points.

% \revanth{To address this xx \ra Should we add hereby referred to as point comparison}
The query point $q_j$ intersects the AABBs of data points $p_1$ and $p_2$. However, this is insufficient to determine if points $p_1$ and $p_2$ are neighbors of $q_j$. To address this, prior work~\cite{zhu2022rtnn} performs a Euclidean distance (or point comparison) test in the intersection shader to identify the exact neighbors within $\epsilon$ distance. 

A key difference between our work and prior work is that while we utilize the ray-AABB intersection test to determine if a point is a potential neighbor, we do not perform the Euclidean distance test. Instead, we use a separate CUDA core-only kernel to refine candidates after searching the BVH tree using RT cores. We define a point $p_r$ as a \emph{candidate} (potential) neighbor for a query point $q_j$ if $q_j$ intersects the AABB of $p_r$.

\section{\ouralg: Ray Tracing Accelerated High Dimensional Vector Similarity Search}\label{sec:ourAlg}

The steps of our proposed \ouralg algorithm are summarized below and described in the associated subsections.

\begin{enumerate}

    \item Select a good combination of three RT indexing dimensions from $d$ dimensions (Section~\ref{sec:identifyDims}).
    \item Identify and group nearby points using a kd-tree and represent the $\epsilon$-neighborhood of nearby points as a primitive (Section~\ref{sec:coarseIndex}).
    \item Rearrange $D$ in memory to improve caching and employ a result mask to store the result set (Section~\ref{sec:reorderPoints}).
    \item Employ a two-pass ray tracing batching scheme to address limited GPU global memory (Section~\ref{sec:twoPassApproach}).
    \item Refine candidate neighbor points by tiling query and candidate neighbor points in shared memory (Section~\ref{sec:refine}).
    \item Enable efficient result set transfers from device to host (Section~\ref{sec:copyResults}).
\end{enumerate}

\subsection{Selecting Indexing Dimensions on RT Cores}\label{sec:identifyDims}

Recall that RT cores are limited to three dimensions (Section~\ref{sec:problemStatement}). To address this limitation, we utilize RT cores to index on three dimensions and identify candidate neighbor points for all queries, $q_j\in Q$. Recent works in the literature have demonstrated that dimensions with higher variance prune the data space more efficiently than those with relatively lower variance~\cite{coss, gdsJoin, gdbod}. This is because dimensions with higher variance tend to have more spread between the coordinates of data points. We refer to a good combination of dimensions for indexing as indexing dimensions. These indexing dimensions are identified by computing the statistical variance of each dimension in the dataset and reordering the coordinates of the data points in memory such that dimensions with higher variance are placed before those with lower variance. We then use the first three dimensions as the indexing dimensions for RT cores. Reordering the dataset also helps in the early termination of the distance computations during the refinement of candidate points using CUDA cores (Section~\ref{sec:refine}).

\subsection{Identifying and Grouping Nearby Points}\label{sec:coarseIndex}

\renewcommand{\arraystretch}{0.5} % Default value: 1
\begin{figure}[!t] 
  \centering
  \begin{tabularx}{\columnwidth}{Y}
   % \documentclass[tikz,border=10pt]{standalone}
% \usepackage{tikz}

% \begin{document}

\begin{tikzpicture}[scale=0.6]

    % \node[above] at (-4,1) {\small {(a)}};

    % Point p_1
    % \draw[thin] (0,0.5) circle (1);
    %  \fill (0,0.5) circle (1.5pt);
    %  \node[below] at (0,0.5) {\small $p_2$};

    %  \draw[thin] (0.65,1.35) circle (1);
    %  \fill (0.65,1.35) circle (1.5pt);
    %  \node[above] at (0.65,1.35) {\small $p_0$};

    %   \draw[thin] (-1.05,1.15) circle (1);
    %  \fill (-1.05,1.15) circle (1.5pt);
    %  \node[below] at (-1.05,1.15) {\small $p_4$};

    %  \draw[thin] (4.05, 0.35) circle (1);
    %  \fill (4.05,0.35) circle (1.5pt);
    %  \node[below] at (4.05,0.35) {\small $p_1$};

    %   \draw[thin] (5.25, 0.35) circle (1);
    %  \fill (5.25,0.35) circle (1.5pt);
    %  \node[below] at (5.25,0.35) {\small $p_3$};

    %   \draw[thin] (1, -1.05) circle (1);
    %  \fill (1,-1.05) circle (1.5pt);
    %  \node[below] at (1, -1.05) {\small $p_5$};
     
    %  %For marking epsilon
    %  \draw[thick, dotted] (1,-1.05) -- (2.0, -1.24);
    % \node[above] at (1.5, -1.15) {\footnotesize$\epsilon$};

    \fill[blue] (0.5,-0.33) circle (2pt);
    \node at (-0.1, 0) {\small \color{blue} $q_j$};

     % Point p_1
     \draw[thin] (0.3,0.55) circle (1);
     \fill (0.3,0.55) circle (1.5pt);
     \node[above] at (0.3,0.55) {\small $p_2$};

     \draw[thin] (0.65,1.35) circle (1);
     \fill (0.65,1.35) circle (1.5pt);
     \node[above] at (0.65,1.35) {\small $p_0$};

     \draw[thin] (-1.05,1.15) circle (1);
     \fill (-1.05,1.15) circle (1.5pt);
     \node[above] at (-1.05,1.15) {\small $p_4$};

     \draw[thin] (4.35, 0.35) circle (1);
     \fill (4.35,0.35) circle (1.5pt);
     \node[below left] at (4.35,0.35) {\small $p_1$};

     \draw[thin] (5.25, 0.35) circle (1);
     \fill (5.25,0.35) circle (1.5pt);
     \node[below right] at (5.25,0.35) {\small $p_3$};

     \draw[thin] (1, -1.05) circle (1);
     \fill (1,-1.05) circle (1.5pt);
     \node[left] at (1, -1.05) {\small $p_5$};

        % For p_6
     \draw[thin] (1.9, -1.05) circle (1);
     \fill (1.9,-1.05) circle (1.5pt);
     \node[right] at (1.9, -1.05) {\small $p_6$};

     %For marking epsilon
     \draw[thick, dotted] (4.35, 0.35) -- (3.35, 0.35);
     \node[above] at (4, 0.35) {\footnotesize$\epsilon$};

\end{tikzpicture}

% \end{document}
 \\
   \footnotesize \textbf{\makecell[l]{(a) Points in the dataset $D$ where the $\epsilon$-neighborhood  of each point is \\ represented as a primitive. Query point $q_j$ is highlighted in blue.}} \\
   \\ 
    % \documentclass[tikz,border=10pt]{standalone}
% \usepackage{tikz}

% \begin{document}

\begin{tikzpicture}[scale=0.6]

% \revanth{put epsilon to represent search distance}
    % \node[above] at (-1.5,3) {\small {(b)}};

    % Point p_1
    % \draw[thin] (0,0.5) circle (1);
    %  \fill (0,0.5) circle (1.5pt);
    %  \node[below] at (0,0.5) {\small $p_2$};

    %  \draw[thin] (0.65,1.35) circle (1);
    %  \fill (0.65,1.35) circle (1.5pt);
    %  \node[above] at (0.65,1.35) {\small $p_0$};

    %   \draw[thin] (-1.05,1.15) circle (1);
    %  \fill (-1.05,1.15) circle (1.5pt);
    %  \node[below] at (-1.05,1.15) {\small $p_4$};

    %  \draw[thin] (4.05, 0.35) circle (1);
    %  \fill (4.05,0.35) circle (1.5pt);
    %  \node[below] at (4.05,0.35) {\small $p_1$};

    %   \draw[thin] (5.25, 0.35) circle (1);
    %  \fill (5.25,0.35) circle (1.5pt);
    %  \node[below] at (5.25,0.35) {\small $p_3$};

    %   \draw[thin] (1, -1.05) circle (1);
    %  \fill (1,-1.05) circle (1.5pt);
    %  \node[below] at (1, -1.05) {\small $p_5$};

    %   \draw[thick, dotted] (1,-1.05) -- (2.0, -1.24);
    % \node[above] at (1.5, -1.15) {\footnotesize$\epsilon$};

    % % MBC for three points
    %  \draw[thick, dotted] (-0.18, 1.15) circle (1.9);
    %  \node[below] at (0.9, 3.3) { $s_0$};

    %  % MBC for two points
    %  \draw[thick, dotted] (4.64, 0.35) circle (1.65);
    %   \node[above] at (5.5, 1.8) { $s_1$};

    %  %MBC for p6 would intersect
    %  \node[above] at (1.7, -0.26) {$s_2$};
     
    %   \draw[thick, dotted] (1, -1.05) circle (1);

    % \fill (0.4,-0.55) circle (1.5pt);
    % \node[below] at (0.4,-0.55) {\small $q_j$};

    \fill[blue] (0.5,-0.33) circle (2pt);
    \node at (-0.1, 0) {\small \color{blue} $q_j$};

     \draw[thin] (0.3,0.55) circle (1);
     \fill (0.3,0.55) circle (1.5pt);
     \node[above] at (0.3,0.55) {\small $p_2$};

     \draw[thin] (0.65,1.35) circle (1);
     \fill (0.65,1.35) circle (1.5pt);
     \node[above] at (0.65,1.35) {\small $p_0$};

      \draw[thin] (-1.05,1.15) circle (1);
     \fill (-1.05,1.15) circle (1.5pt);
     \node[above] at (-1.05,1.15) {\small $p_4$};

     \draw[thin] (4.35, 0.35) circle (1);
     \fill (4.35,0.35) circle (1.5pt);
     \node[below left] at (4.35,0.35) {\small $p_1$};

      \draw[thin] (5.25, 0.35) circle (1);
     \fill (5.25,0.35) circle (1.5pt);
     \node[below right] at (5.25,0.35) {\small $p_3$};

      \draw[thin] (1, -1.05) circle (1);
     \fill (1,-1.05) circle (1.5pt);
     \node[left] at (1, -1.05) {\small $p_5$};

      % For p_6
     \draw[thin] (1.9, -1.05) circle (1);
     \fill (1.9,-1.05) circle (1.5pt);
     \node[right] at (1.9, -1.05) {\small $p_6$};

     %MBC for p_5 and p_6
      \draw[ thick, dotted] (1.45, -1.05) circle (1.5);
     \node[below] at (2.45, 0.95) { $s_2$};

      \draw[ thick, dotted] (4.35, 0.35) -- (3.35, 0.35);
     \node[above] at (4, 0.35) {\footnotesize$\epsilon$};

    % MBC for three points
     \draw[ thick, dotted] (-0.18, 1.15) circle (1.9);
     \node[below] at (1.2, 3.4) { $s_0$};

     % MBC for two points
     \draw[thick, dotted] (4.8, 0.35) circle (1.5);
      \node[above] at (5.8, 1.45) { $s_1$};

     %MBC for p6 would intersect
     % \node[above] at (1.7, -0.26) {$s_2$};
     
     %  \draw[thick, dotted] (1, -1.05) circle (1);

    % Separate query point

\end{tikzpicture}

% \end{document}
 \\
    \footnotesize \textbf{\makecell[l]{(b) Grouping and representing the $\epsilon$-neighborhood of nearby \\points as a primitive $s$. }}\\
    \\ 
    % \documentclass[tikz,border=10pt]{standalone}
% \usetikzlibrary{decorations.pathreplacing}
% \usepackage{tikz}
% \usetikzlibrary {arrows.meta}

% \begin{document}

\begin{tikzpicture}[scale=0.5]

    % \node[above] at (-6,3) {\small {(c)}};
     
    \foreach \x in {0,1,2,3,4,5,6}
    {
        \draw[thick] (\x,0) rectangle (\x+1,1);
    }

    \node at (-1, 2.5) {$D$};

    \node at (-1, 0.5) {$D^{\prime}$};
    
    \node at (0.5,0.5) {$p_4$};
    \node at (1.5,0.5) {$p_0$};
    \node at (2.5,0.5) {$p_2$};
    \node at (3.5,0.5) {$p_5$};
    \node at (4.5,0.5) {$p_6$};
    \node at (5.5,0.5) {$p_1$};
    \node at (6.5,0.5) {$p_3$};

    \foreach \x in {0,1,2,3,4,5, 6}
    {
        \draw[thick] (\x,2) rectangle (\x+1,3);
        \node at (\x+0.5,2.5) {$p_{\x}$};
    }

% \revanth{Put D to reference Dataset}
    \draw [decorate,decoration={brace,amplitude=6pt}]
    (2.9,-0.1) -- (0.1,-0.1)
    node[midway, yshift=-12pt] {$s_0$};

    \draw [decorate,decoration={brace,amplitude=6pt}]
    (4.9,-0.1) -- (3.1,-0.1)
    node[midway, yshift=-12pt] {$s_2$};

    \draw [decorate,decoration={brace,amplitude=6pt}]
    (6.9,-0.1) -- (5.1,-0.1)
    node[midway, yshift=-12pt] {$s_1$};

\draw[arrows = {-Stealth}] (0.5,2) -- (1.5,1); 
\draw[arrows = {-Stealth}] (1.5,2) -- (5.5,1); 
\draw[arrows = {-Stealth}] (2.5,2) -- (2.5,1); 
\draw[arrows = {-Stealth}] (3.5,2) -- (6.5,1); 
\draw[arrows = {-Stealth}] (4.5,2) -- (0.5,1); 
\draw[arrows = {-Stealth}] (5.5,2) -- (3.5,1); 
\draw[arrows = {-Stealth}] (6.5,2) -- (4.5,1); 
    
\end{tikzpicture}

% \end{document}
 \\
    \footnotesize \textbf{\makecell[l]{(c) All points in each primitive $s$ are rearranged contiguously in $D^\prime$.}} \\
    \\ 
    % \documentclass[tikz,border=10pt]{standalone}
% \usetikzlibrary{decorations.pathreplacing}
% \usepackage{tikz}
% \usetikzlibrary {arrows.meta}

% \begin{document}

\begin{tikzpicture}[scale=0.5]

    % \node at (-1,2) {(c)};

    % For self join
    \foreach \x in {0,1,2,3,4,5,6}
    {
        \draw[thick] (\x,0) rectangle (\x+1,1);
    }

    % For semi join
    \foreach \x in {9.5}
    {
        \draw[thick] (\x,0) rectangle (\x+1,1);
    }

    \node at (10, 0.5) {$q_j$};

    \draw[thick] (8.66,-2) rectangle (10,-1);
    \draw[thick] (10,-2) rectangle (11.33,-1);

     \node at (9.33, -1.5) {$s_0$};
    \node at (10.67, -1.5) {$s_2$};

    \foreach \x in {8, 9, 10, 11, 12}
    {
        \draw[thick] (\x,-4) rectangle (\x+1,-3);
    }

    \node at (8.5,-3.5) {$p_4$};
    \node at (9.5,-3.5) {$p_0$};
    \node at (10.5,-3.5) {$p_2$};
    \node at (11.5,-3.5) {$p_5$};
    \node at (12.5,-3.5) {$p_6$};

     \draw[arrows = {-Stealth}, dash dot] (10,0) -- (9.33,-1);
    \draw[arrows = {-Stealth}, dash dot] (10,0) -- (10.67,-1);

    \draw[arrows = {-Stealth}] (9.33,-2) -- (8,-3);
    \draw[arrows = {-Stealth}] (9.33,-2) -- (11,-3);

    \draw[arrows = {-Stealth}] (10.67,-2) -- (11,-3);
    \draw[arrows = {-Stealth}] (10.67,-2) -- (13,-3);

     \draw [decorate,decoration={brace,amplitude=6pt}]
    (10.9,-4.1) -- (8.1,-4.1)
    node[midway, yshift=-12pt] {$s_0$};

    \draw [decorate,decoration={brace,amplitude=6pt}]
    (12.9,-4.1) -- (11.1,-4.1)
    node[midway, yshift=-12pt] {$s_2$};

    %===========================

    \node at (0.5,1.5) {$p_4$};
    \node at (1.5,1.5) {$p_0$};
    \node at (2.5,1.5) {$p_2$};
    \node at (3.5,1.5) {$p_5$};
    \node at (4.5,1.5) {$p_6$};
    \node at (5.5,1.5) {$p_1$};
    \node at (6.5,1.5) {$p_3$};

    \node at (0.5,0.5) {$q_4$};
    \node at (1.5,0.5) {$q_0$};
    \node at (2.5,0.5) {$q_2$};
    \node at (3.5,0.5) {$q_5$};
    \node at (4.5,0.5) {$q_6$};
    \node at (5.5,0.5) {$q_1$};
    \node at (6.5,0.5) {$q_3$};

    \node at (-1.5, 0.5) {$Q = D^\prime$};
    \node at (8.5, 0.5) {$Q$};

    % \node at (-1, -3.5) {$D^{\prime}$};

    \node at (-1, -1.5) {$S$};

    \node at (-1, -3.5) {$D^{\prime}$};

    % \draw[thick] (1,-2) rectangle (2.33,-1);
    % \node at (1.67, -1.5) {$s_0$};

    % \draw[thick] (2.33,-2) rectangle (3.66,-1);
    % \node at (3, -1.5) {$s_1$};
    
    % \draw[thick] (3.66,-2) rectangle (4.99,-1);
    % \node at (4.33, -1.5) {$s_2$};

    % consistent width
    \def\w{1.33}
    
    \draw[thick] (1.5,-2) rectangle ++(\w,1);
    \node at (1.5+0.5*\w,-1.5) {$s_0$};
    
    \draw[thick] (1.5+\w,-2) rectangle ++(\w,1);
    \node at (1.5+1.5*\w,-1.5) {$s_2$};
    
    \draw[thick] (1.5+2*\w,-2) rectangle ++(\w,1);
    \node at (1.5+2.5*\w,-1.5) {$s_1$};

    \draw[arrows = {-Stealth}, dash dot] (0.5,0) -- (2.33,-1); 
    \draw[arrows = {-Stealth}, dash dot] (1.5,0) -- (2.33, -1); 
    \draw[arrows = {-Stealth}, dash dot] (2.5,0) -- (2.33,-1); 
    % \draw[arrows = {-Stealth}, dash dot] (5.5,0) -- (1.67,-1); 

    % \draw[arrows = {-Stealth}, dash dot] (0.5,0) -- (4.33,-1); 
    % \draw[arrows = {-Stealth}, dash dot] (1.5,0) -- (4.33,-1); 
    % \draw[arrows = {-Stealth}, dash dot] (2.5,0) -- (4.33,-1); 
    
     \draw[arrows = {-Stealth}, dash dot] (3.5,0) -- (3.5,-1); 
     \draw[arrows = {-Stealth}, dash dot] (4.5,0) -- (3.5,-1); 
      
      \draw[arrows = {-Stealth}, dash dot] (5.5,0) -- (4.67,-1); 
      \draw[arrows = {-Stealth}, dash dot] (6.5,0) -- (4.67,-1);

    \foreach \x in {0,1,2,3,4,5,6}
    {
        \draw[thick] (\x,-4) rectangle (\x+1,-3);
    }

     \node at (0.5,-3.5) {$p_4$};
    \node at (1.5,-3.5) {$p_0$};
    \node at (2.5,-3.5) {$p_2$};
    \node at (3.5,-3.5) {$p_5$};
    \node at (4.5,-3.5) {$p_6$};
    \node at (5.5,-3.5) {$p_1$};
    \node at (6.5,-3.5) {$p_3$};

    % \node at (-1.5,-3.5) {$p_4$};
    % \node at (-0.5,-3.5) {$p_0$};
    % \node at (0.5,-3.5) {$p_2$};
    % \node at (1.5,-3.5) {$p_5$};
    % \node at (2.5,-3.5) {$p_1$};
    % \node at (3.5,-3.5) {$p_3$};
    %  \node at (4.5,-3.5) {$p_4$};
    % \node at (5.5,-3.5) {$p_0$};
    % \node at (6.5,-3.5) {$p_2$};
    % \node at (7.5,-3.5) {$p_5$};

    \draw [decorate,decoration={brace,amplitude=6pt}]
    (2.9,-4.1) -- (0.1,-4.1)
    node[midway, yshift=-12pt] {$s_0$};

    \draw [decorate,decoration={brace,amplitude=6pt}]
    (4.9,-4.1) -- (3.1,-4.1)
    node[midway, yshift=-12pt] {$s_2$};

    \draw [decorate,decoration={brace,amplitude=6pt}]
    (6.9,-4.1) -- (5.1,-4.1)
    node[midway, yshift=-12pt] {$s_1$};

    % Draw lines
     \draw[arrows = {-Stealth}] (1.67,-2) -- (0,-3); 
     \draw[arrows = {-Stealth}] (1.67,-2) -- (3,-3); 

     \draw[arrows = {-Stealth}] (3,-2) -- (3,-3); 
     \draw[arrows = {-Stealth}] (3,-2) -- (5,-3); 

     \draw[arrows = {-Stealth}] (4.33,-2) -- (5,-3); 
     \draw[arrows = {-Stealth}] (4.33,-2) -- (7,-3); 

    %===========================

    \node at (3.5,-6) {\small Self-join};
    \node at (10.7,-6) {\small Semi-join};

\end{tikzpicture}

% \end{document}
 \\ 
    \\ 
    \footnotesize \textbf{\makecell[l]{(d) Queries intersecting the same primitive are stored contiguously, \\with primitives having the highest number of intersections stored first.}}
 \end{tabularx}
 % \revanth{}

 \caption{Figure illustrating our \ouralg algorithm (see text for details).}
 \label{fig:master}
\end{figure}
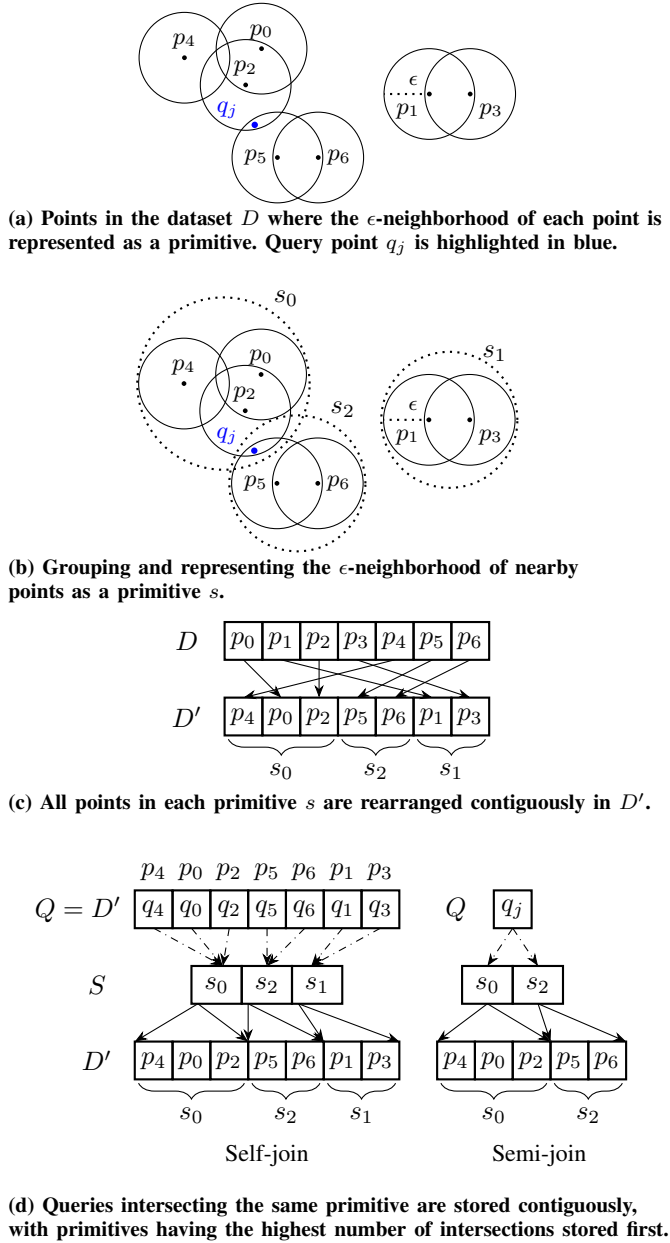

\renewcommand{\arraystretch}{1} % Default value: 1

We represent the $\epsilon$-neighborhood of point $p_i \in D$ using the first 3 dimensions and represent $p_i$ as a primitive (sphere) $s_i$ in the \optix scene with center at $\{x_i^1, x_i^2, x_i^3\}$. For each point $p_i \in D$, we map a primitive $s_i$ and store it in an array $S$. 

\noindent\textbf{Traditional Approach:} Prior work that indexes using RT cores involves constructing the BVH tree using the primitives in the array $S$. For each query, a ray is generated following the steps outlined in Section~\ref{sec:rangeSearches} to identify neighbors (Figure~\ref{fig:master}(a)). However, this approach might not yield good performance as RT searches using BVH tree traversals suffer from a high number of ray-primitive intersections on large datasets, which degrades performance.

\noindent\textbf{Our Approach:} To address this challenge, as shown in Figure~\ref{fig:master}(b), we identify nearby points in the data space and instead of representing the $\epsilon$-neighborhood of each point in the dataset as an individual primitive in the scene, we group $\epsilon$-neighborhood of nearby points in the data space together. Specifically, we identify a bounding sphere that encompasses the $\epsilon$-neighborhood of nearby points. These bounding spheres are inserted as primitives in the \optix scene, and to maintain consistent terminology with the ray-tracing literature, we refer to these bounding spheres as primitives. A query $q_j$ intersecting the AABB of the bounding sphere during the BVH tree traversal implies that all of the individual data points within the primitive are candidate neighbors for the query.

We find that perhaps surprisingly, grouping the $\epsilon$-neighborhood of nearby points in the data space together is a good task for an indexing method. This is because indexes partition the space and in doing so implicitly group nearby points which we then use to create bounding spheres.

We construct a balanced kd-tree of height $h_{kd}$ on the GPU as we find that it is a good index for this purpose because it has a tunable parameter to control the number of levels/height, and a balanced kd-tree ensures that each leaf node contains roughly the same number of points.\footnote{Each leaf node has roughly  the same number of points because some leaf nodes will contain more points than others if $|D|$ is not a power of two.} This yields uniform number of points within each bounding sphere. We exploit this property to design an efficient batching scheme with minimal load imbalance in Section~\ref{sec:batching}. We denote $|L|$ as the number of points in each kd-tree leaf node which are stored in each bounding sphere, and denote \emph{primitive points} as points within a bounding sphere. A ray mapped to a query intersecting the AABB of the bounding sphere implies that all the points contained within are candidate neighbors of the query point that will need to be refined by the CUDA kernel.

There is a trade-off regarding the height of the kd-tree. A short kd-tree with few levels will have few leaf nodes and thus fewer bounding spheres. This implies that there will be many candidate points that will need to be refined as each sphere will map to a large number of points in the dataset. In contrast, a kd-tree with more levels will generate a larger number of bounding spheres which improves pruning efficiency, but it may also result in higher tree traversal costs for large datasets. Hence, selecting a good kd-tree height is critical for the overall performance of the algorithm. Although other indexes including Morton, Hilbert, and Ball trees do not provide all the advantages of a kd-tree, we evaluated them and found that the kd-tree provides better overall performance.

Figure~\ref{fig:master}(a) illustrates a sample two-dimensional data space for the dataset $D$, where circles denote the $\epsilon$-neighborhood  of each point with search distance $\epsilon$. Here, $p_0$ is a neighbor of $p_2$ because $p_0$ lies within the $\epsilon$-neighborhood  of $p_2$ and vice versa. Similarly, points $p_1$ and $p_3$, and $p_5$ and $p_6$ lie within each other’s $\epsilon$-neighborhood  and are therefore neighbors. This represents the traditional approach to DSS using RT cores, as described in Section~\ref{sec:rangeSearches}.

Representing each point as a primitive results in $|D|$ primitives in the \optix scene and leads to a large number of ray–primitive intersections, particularly for large datasets, resulting in poor performance. Figure~\ref{fig:master}(b) illustrates our proposed approach, in which nearby points are grouped together and the $\epsilon$-neighborhood  of each group is represented as a single primitive. As shown in Figure~\ref{fig:master}(b), points $p_4$, $p_0$, and $p_2$ are close to each other, and we represent the $\epsilon$-neighborhood  encompassing these three points as a single primitive $s_0$. Similarly, points $p_1$ and $p_3$, and $p_5$ and $p_6$ are grouped together and are represented as primitive $s_1$ and $s_2$, respectively.

\subsection{Rearranging the Points Within the Dataset}\label{sec:reorderPoints}

For a given query, we traverse the BVH tree using RT cores to identify candidate neighbor points. However, these points are typically stored close to each other in global memory, leading to uncoalesced accesses during further refinement (Section~\ref{sec:refine}). Uncoalesced memory accesses result in a large number of memory transactions, poor cache hits, and ultimately, poor performance overall.

To address this problem, we rearrange $p_i \in D$ so that all points belonging to the same bounding sphere are stored contiguously in memory. Additionally, a query mapped to a ray often intersects with multiple bounding spheres. To further optimize memory access patterns, we order the group of points between two primitives using Morton ordering~\cite{morton}. We assume that primitives closer to each other in data space are more likely to be intersected by a given query, and accessing points close to each other in global memory promotes coalesced memory accesses and improves performance.

Figure~\ref{fig:master}(c) illustrates the dataset before and after rearranging. Here, $D$ represents the original layout of points in the dataset, whereas $D^{\prime}$ represents the dataset after rearranging. As shown in Figure~\ref{fig:master}(b), we represent points $p_4$, $p_0$, and $p_2$ as primitive $s_0$; points $p_1$ and $p_3$ as primitive $s_1$; and points $p_5$ and $p_6$ as primitive $s_2$. We then rearrange the points in dataset $D$ such that points within primitive $s_0$ ($p_4$, $p_0$, and $p_2$) are contiguous in memory. Similarly, points within primitive $s_1$ ($p_1$ and $p_3$) and $s_2$ ($p_5$ and $p_6$) are also contiguous. The points belonging to primitives $s_0$ and $s_2$ are stored together as a query intersecting the primitive  $s_0$ is highly likely to intersect $s_2$ as well, and vice versa. This rearranging enables coalesced memory access patterns during candidate refinement.

\subsection{Batching Queries: Two Pass Ray Tracing Approach}\label{sec:twoPassApproach}

GPUs typically have limited global memory compared to CPU main memory, and estimating the memory required to store the results for a given batch of queries is pivotal for saturating GPU resources (Section~\ref{Sec:intro}). To ensure that we saturate resources while accommodating large result set sizes that may exceed GPU global memory capacity, for each query, we traverse the BVH tree using RT cores to count the total number of primitives intersected by all queries in the scene. We also count the number of queries intersected by each primitive. We refer to this as the First Intersection Test (FIT).

Recall that all primitives in the \optix scene represent roughly the same number of points (Section~\ref{sec:coarseIndex}). Hence, the memory required to refine candidate neighbor points for each ray–primitive intersection is approximately the same. We use the FIT to determine the number of primitives intersected by a given query point $q_j$ and identify the maximum number of ray–primitive intersections ($I_{max}$) that can be processed in a single GPU batch by referencing the available global memory using \emph{cudaMemGetInfo}~\cite{cudamemory}.

After rearranging the dataset, we process queries in the same order. This order is chosen because consecutive queries often belong to the same leaf node and have similar primitive intersections and traverse the same BVH tree path, which helps mitigate ray divergence~\cite{zhu2022rtnn}. The query set $Q^\prime \subseteq Q$ is constructed using a window approach, in which queries are accumulated until the total number of intersections are less than or equal to the previously identified $ I_{max}$.

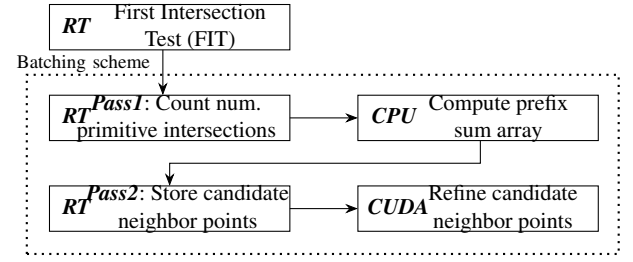
\begin{figure}[!t] 

  \centering
  \begin{tabularx}{\columnwidth}{Y}
   % \documentclass[tikz,border=5pt]{standalone}
% \usetikzlibrary{shapes.geometric, arrows}

% % Styles
% \tikzstyle{process} = [rectangle, minimum height=1.25cm, text width=2.5cm, align=center, draw=black]
% \tikzstyle{arrow} = [,->,>=stealth]

% \begin{document}

\begin{tikzpicture}[scale=0.6, transform shape]

\draw[] (0,0) rectangle (5.3,1);
\node at (3.1,0.5) {\Large{\makecell{First Intersection\\ Test (FIT)}}};
\node at (0.6,0.5) {\textit{\textbf{\Large{{\makecell{RT}}}}}};
\draw[arrows={-Stealth}] (2.5,0) -- (2.5,-1);

\draw[] (0,-2) rectangle (5.3,-1);
\node at (2.8,-1.5) {\Large{\makecell{\textbf{\emph{Pass1}}: Count num.\\primitive intersections}}};
\node at (0.6,-1.5) {\textit{\textbf{\Large{{\makecell{RT}}}}}};
\draw[arrows={-Stealth}] (5.3,-1.5) -- (6.8,-1.5);

\draw[] (6.8,-2) rectangle (12.1,-1);
\node at (9.9,-1.5) {\Large{\makecell{Compute prefix \\ sum array}}};
\node at (7.6,-1.5) {\textit{\textbf{\Large{{\makecell{CPU}}}}}};
\draw[arrows={-Stealth}] (9.5,-2) -- (9.5, -2.5) -- (2.65, -2.5) -- (2.65,-3);

% {\Large{\makecell{Compute prefix\\ sum array}}}

\draw[] (6.8,-4) rectangle (12.1,-3);
\node at (10,-3.5) {\Large{\makecell{Refine candidate\\neighbor points}}};
\node at (7.7,-3.5) {\textit{\textbf{\Large{{\makecell{CUDA}}}}}};
\draw[arrows={-Stealth}] (5.3,-3.5) -- (6.8,-3.5);

\draw[] (0,-4) rectangle (5.3,-3);
\node at (0.6,-3.5) {\textit{\textbf{\Large{{\makecell{RT}}}}}};
\node at (3.05,-3.5) {\Large{\makecell{\textbf{\emph{Pass2}}: Store candidate\\neighbor points}}};

\node at (0.75, -0.3) {\large Batching scheme};
\draw[, dotted, thick] (-0.5,-4.5) rectangle (12.55,-0.55);
% \revanth{Change order}

\end{tikzpicture}

% \end{document}
 \\
 \end{tabularx}
 \caption{Two-pass ray tracing approach used in \ouralg.}
 \label{fig:flowChart2}
\end{figure}

 Figure~\ref{fig:flowChart2} illustrates our \emph{two-pass} ray tracing approach employed by our batching scheme, described as follows:

     \noindent \textbf{RT Pass 1:} For a given query set $Q^\prime \subseteq Q$, we traverse the BVH tree to determine the number of primitive intersections for each query $q_j \in Q^\prime$ and the number of queries intersected by each primitive in the scene.\footnote{The number of primitives intersected by each query can be obtained from the FIT. However, the number of queries intersecting each primitive changes across batches as $Q^\prime$ varies across batches.} Using this, we build a prefix sum array on the CPU to indicate the positions to begin writing the query index intersected for each primitive in the scene.

     \noindent \textbf{RT Pass 2:} For each query $q_j$, we generate a ray and store the indices of the primitives that intersect with the ray, using the prefix sum array constructed in the previous step.

This \emph{two-pass} approach outlined in Figure~\ref{fig:flowChart2} offers several advantages over the conventional method of storing the results using key-value pairs: ($i$) It eliminates the need for global atomic updates to identify the next position to write when storing elements in the result set. ($ii$) It reduces the memory usage by half as we store only the query index, instead of both query and candidate index, as all the queries intersected by a primitive are stored together. 

Lastly, the prefix sum array is constructed such that primitives with more intersections are processed first which mitigates load imbalance towards the end of the execution.

Now that we have described batching using the two-pass RT approach, we illustrate how candidate neighbor points are identified using RT cores in Figure~\ref{fig:master}(d). Here, $Q = \{q_4, q_0, q_2, q_5, q_6, q_1, q_3\}$ represents the query set for the self-join scenario (query set is the same as the dataset) defined in Section~\ref{sec:problemStatement}, which includes all points in dataset $D^\prime$. As shown in Figure~\ref{fig:master}(b), we use primitives $s_0$, $s_1$, and $s_2$ to construct the BVH tree for identifying candidate neighbors for points in the query set $Q$. Points $p_4$, $p_0$, and $p_2$ intersect the axis-aligned bounding box (AABB) of primitive $s_0$, as they lie within that primitive. Similarly, points $p_5$ and $p_6$, and $p_1$ and $p_3$ intersect the AABB of primitive $s_2$ and $s_1$, respectively. All the points within the primitive are candidate neighbors for the query point. For a query point $q_j$ in the semi-join scenario (query set is distinct from the dataset), $q_j$ intersects the AABB of primitives $s_0$ and $s_2$. Consequently, all points contained within primitives $s_0$ and $s_2$ are considered candidate neighbor points for the query $q_j$.

\subsection{Candidate Refinement on CUDA Cores}\label{sec:refine}

The prior subsections largely discussed RT cores, these include selecting indexing dimensions,  constructing bounding spheres, rearranging point data, and executing batches. The RT core phase in a given batch records the candidate data, which are those points that may be within $\epsilon$ search distance of the batch of queries, $Q^\prime$.

Now that the candidates have been established in global memory, we execute a refinement kernel that uses CUDA cores. A straightforward approach is to perform Euclidean distance calculations between each query point and each point within a bounding sphere by reading point coordinate data directly from global memory. However, this approach is inefficient due to (slow) off-chip global memory accesses. To improve performance, we present the following steps: ($i$) We assign a thread block to identify neighbors for all queries that intersect a given primitive. We allocate shared memory for each thread block. ($ii$)  We divide the shared memory into two sections: one for loading query points and the other for loading primitive points from global memory. ($iii$) All threads within a block cooperate to tile the primitive points into shared memory. All the coordinates of a given point in $D$ are stored contiguously. However, when copying into shared memory from global memory, we store the coordinates of all points belonging to the same dimension contiguously (i.e., thus transposing the data). We tile the query points into shared memory using the same approach as above. ($iv$) We refine candidate neighbor points by accessing shared memory and performing Euclidean distance calculations (point comparisons).

The shared memory tiling enables data reuse and mitigates slow global memory accesses. We show that it is more efficient than directly reading the point data from global memory.

\subsection{Result Representation \& Data Transfers}\label{sec:copyResults}

% \mike{You have text that says "Storing the result set as key-value pairs (\label{sec:resultRep})" which is wrong. It needs to be commented because you're redefining the label and making Section II-A1 be relabeled to Section IV-F. }
Storing the result set as key-value pairs (Section~\ref{sec:resultRep}) requires a significant amount of space. This limits the number of queries that can be processed in a single batch and also impacts the time required to copy the results from the device (GPU) to the host (CPU).

Recall that all queries that intersect with a given primitive are stored together (Section~\ref{sec:twoPassApproach}). We use a result mask where the result of a comparison between a query and point within the primitive is represented using a single bit (0 or 1). If there are $|Q^\prime|$ queries in a given batch and $|S^\prime|$ primitives intersected across all queries in this batch, we only require $|Q^\prime| \cdot |S^\prime| \cdot |L|$ bits to record whether a query point is within $\epsilon$ of the candidate points. Using the result mask significantly reduces the memory required to store the result set and enables more queries to be processed in a single batch.

\begin{figure}[!t]

\centering
         \begin{tikzpicture}[scale=0.4]

    \foreach \x in {0,1,2,3,4,5,6,7,8,9,10,11,12,13}
    {
        \draw[thick] (\x,0) rectangle (\x+1,1);
         \node at ( \x+0.5, 1.5) {\x};
    }
    
    \node at (-1.5, 0.5) {$R$};
     \node at (0.5,0.5) {0};
     \node at (1.5,0.5) {1};
     \node at (2.5,0.5) {0};
     \node at (3.5,0.5) {0};
     \node at (4.5,0.5) {0};
     \node at (5.5,0.5) {1};
     \node at (6.5,0.5) {0};
     \node at (7.5,0.5) {0};
     \node at (8.5,0.5) {1};
     \node at (9.5,0.5) {1};
     \node at (10.5,0.5) {0};
     \node at (11.5,0.5) {0};
     \node at (12.5,0.5) {1};
     \node at (13.5,0.5) {0};

    \foreach \x in {0,1,2,3,4}
    {
        \draw[thick] (\x + 4.0, -3) rectangle (\x+5,-2);
    }

    \node at (2.5,-2.5) {{$R^{\prime}$}}; 
    \node at (4.5,-2.5) {1};
    \node at (5.5,-2.5) {5};
    \node at (6.5,-2.5) {8};
    \node at (7.5,-2.5) {9};
    \node at (8.5,-2.5) {12};

     \draw[arrows = {-Stealth}] (6.5,0) -- (6.5,-2); 
     
\end{tikzpicture}     
    \caption{Stream compaction employed in \ouralg. Here, $R$ and $R^\prime$ denote the result set before and after compression.}
   \label{fig:runLengthEncoding}
\end{figure}
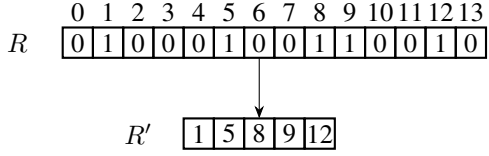

% \revanth{Please read this section.}

% The result mask often contains a long sequence of consecutive zeroes and a few ones, where ones represent neighbors. To minimize the result mask size, we employ stream compaction that compresses the result set into a compact list of set-bit indices which reduces the data size being transferred from device to host. 

\subsubsection{Result Mask Compression}\label{sec:compression}
The result mask often contains a long sequence of consecutive zeroes and a few ones, where ones represent neighbors.
Instead of storing the result of the point comparison for each query and candidate point (i.e., whether they are within $\epsilon$ of each other), we only store the positions of neighbors (ones) and construct a compressed result mask that only contains $|R|$ ones representing the positions of neighbors. This directly reduces the size of result set that is copied from the device to the host. This compression occurs on the GPU after refining candidate neighbor points using CUDA cores.

Figure~\ref{fig:runLengthEncoding} illustrates the stream compaction technique employed in \ouralg. Here, $R$ represents the result mask that stores the outcomes of point comparisons between query points and candidate neighbor points. As shown in the figure, $R$ requires 14 result elements, whereas storing only the indices of point comparisons corresponding to neighbors requires only 5 result elements after stream compaction, denoted by $R^{\prime}$. 

\subsubsection{Pinned Memory for Fast Data Transfers}\label{sec:pinnedMemory}

Another limitation of memory transfers arises from copying data from GPU global memory to pageable memory on the host. Copying data from device to host involves the temporary locking of pages in main memory and then copying to pageable memory. To minimize memory copy time, we allocate a small amount of pinned memory as a staging area and utilize nested parallelism with multiple GPU streams to copy the compressed result mask from the device to the host. This approach decreases the time needed to store results on the host.

\section{Experimental Evaluation}\label{sec:expEval}

\subsection{Experimental Methodology}
The GPU code is written in CUDA and compiled using CUDA version 13.0. The host code is compiled using g++ version 13.3.0 and is parallelized using OpenMP with the O3 optimization flag. The \optix code is written using the \optix Wrappers Library (OWL) with \optix version 7.4.0. OWL is a convenience library built on top of \optix that abstracts many low level details~\cite{OWL}. Experiments are conducted on a platform with an Intel W-2295 CPU with 256 GiB of main memory, equipped with a Quadro RTX 5000 GPU with 16 GiB of global memory.

All response times reported in this section include all time components, including reordering the dataset, grouping points together using a balanced kd-tree, indexing using RT cores, refining candidate neighbor points using CUDA cores, and copying the result set from device to host, and exclude the time to load the dataset from disk. We set a timeout limit of 60,000 seconds for all experiments, and the response time of any experiment exceeding this limit is marked as \timeout. For all experiments, if the corresponding algorithm exceeds the timeout limit, we use the timeout value to compute the lower bound speedup, even though the actual response time may be significantly larger.

\subsection{Datasets}

\begin{table}[t]
\setlength{\tabcolsep}{2pt}

    \caption{The six real-world datasets with their dimensionality ($d$), size ($|D|$), and search distances ($\epsilon$).}
    % \revanth{Update citation}
    \centering
    \footnotesize
    \begin{tabularx}{\columnwidth}{Xrrrrr}
        \hline
         Dataset & $d$ & $|D|$ & $\epsilon_{s}$ & $\epsilon_{m}$ & $\epsilon_{l}$ \\
         \hline
          \susy~\cite{datasetsusyhiggs} & 18 & 5,000,000 & 0.017030 & 0.020780 & 0.025555 \\
          \higgs~\cite{datasetsusyhiggs} & 28 & 11,000,000 & 0.049186 & 0.055580 & 0.063117 \\
          \wave~\cite{datasetwave} & 49 & 287,999 & 0.005400 & 0.007020 & 0.008358 \\
          \bigcross~\cite{ackermann2012streamkm} & 57 & 11,620,300 & 0.013100 & 0.019940 & 0.028100 \\
          \msd~\cite{datasetmsd} & 90 & 515,345 & 0.007600 & 0.009130 & 0.011334\\
          \sift~\cite{datasetSift} & 128 & 10,000,000 & 0.598828 & 0.668554 & 0.748242 \\
         \hline
    \end{tabularx}
    \label{tab:datasets}
\end{table}

Table~\ref{tab:datasets} lists the real-world datasets and search distances ($\epsilon$) employed in the experimental evaluation. The datasets contain 288K--11.6M data points, with a dimensionality $d \in [18, 128]$. We use three search distances, $\epsilon_s$, $\epsilon_m$, and $\epsilon_l$, corresponding to small, medium, and large selectivity levels. Here, the selectivity is defined as $z=\frac{|R| - |D|}{|D|}$, which denotes the average number of neighbors found per query in the query set $Q$~\cite{gdsJoin, coss, donnelly2024multi}. Specifically, $\epsilon_s$, $\epsilon_m$, and $\epsilon_l$ correspond to selectivity values of $2^8$, $2^{10}$, and $2^{12}$, respectively. These selectivity values are consistent with the literature~\cite{donnelly2024multi}.

\subsection{Reference Implementations}\label{sec:refAlg}

We evaluate the performance of \ouralg using the following state-of-the-art (SOTA) GPU-accelerated distance similarity search algorithms.

\noindent{\textbf{\mikealg}}: A similarity search algorithm using grid-based indexing for multi-dimensional data, proposed by Gowanlock and Karsin~\cite{gdsJoin}.\\ 
\noindent{\textbf{\brianalg}}:  A similarity search algorithm using a metric-based indexing method, by Donnelly and Gowanlock~\cite{coss}. \\
\noindent{\textbf{\GTS}}: A similarity search algorithm using a metric-based tree structure for efficient searches, proposed by Zhu et al.~\cite{GTS}. \\
\noindent{\textbf{\pyTorchAlg}}: A similarity search algorithm from the PyTorch3D library by the Facebook AI Research team~\cite{pytorch3d}. \\
\noindent{\textbf{\cuvsalg}}: A brute-force vector similarity search algorithm from the CUDA Vector Search (cuVS) software development kit by NVIDIA~\cite{cuvs}. \\
\noindent{\textbf{\rtnn}}: A RT-accelerated similarity search algorithm. We use this only for $d=3$ dataset performance evaluation. \\

To address the limited global memory issue, \mikealg, \brianalg, and \GTS employ a batching scheme. We implemented a similar approach for \cuvsalg and \pyTorchAlg to handle out-of-memory issues.

% We exclude all RT core-based algorithms outlined in Section~\ref{sec:background} as they are limited to 3 dimensions.

\subsection{kd-tree Height}\label{sec:paramSel}

Recall that the kd-tree height ($h_{kd}$) directly affects the number of primitives in the scene and consequently, the number of ray–primitive intersections. A higher $h_{kd}$ results in more primitives, offering greater pruning but also incurs higher indexing and search costs, whereas a lower $h_{kd}$ provides less pruning and leads to increased work when refining candidate points using the CUDA cores (Section~\ref{sec:coarseIndex}).

\begin{figure}[!t]
\centering
\includegraphics[width=\columnwidth]{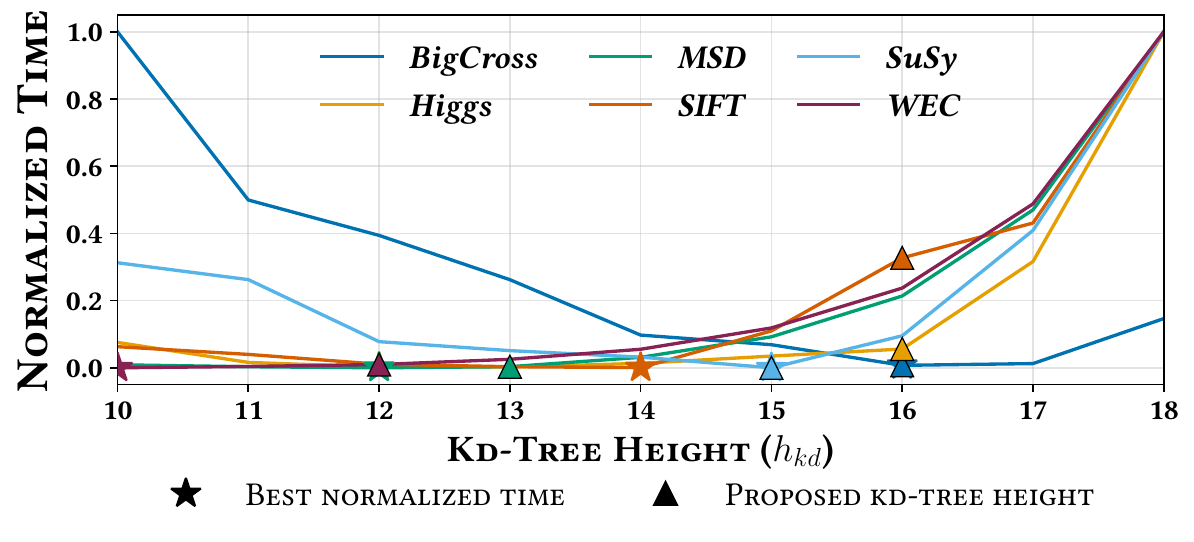}
    % \revanth{Update figure.}
    \caption{Averaged normalized response time for kd-tree heights ($h_{kd}$) in the range [10, 18] across all datasets. The observed best time and the proposed kd-tree height are indicated using star and triangle markers, respectively.}
   \label{fig:kdTreeLevel}
\end{figure}

Figure~\ref{fig:kdTreeLevel} shows the averaged normalized response time across the three search distances ($\epsilon_s$, $\epsilon_m$, and $\epsilon_l$) for kd-tree heights in the range [10, 18]. For the \wave dataset, the best performance is observed at $h_{kd} = 10$, and performance deteriorates as $h_{kd}$ increases. For all other datasets, performance improves with an increase in tree height up to a certain tree height and then degrades later.

We observe that an appropriate kd-tree height $h_{kd}$ depends on the number of points in the dataset, $|D|$. We propose that a good value of $h_{kd}$ can be identified as a function of $|D|$. Specifically, setting $h_{kd} = \lfloor \ln |D| \rfloor$ yields a tree height that performs well across all datasets (triangle markers in Figure~\ref{fig:kdTreeLevel}) and is close to the optimal height  (star markers). Therefore, we use $h_{kd} = \lfloor \ln |D| \rfloor$ in all forthcoming experiments.

\subsection{Impact of Grouping Points Together}\label{sec:expGroupPoints}

\begin{figure}[!t]
\centering
\includegraphics[width=\columnwidth]{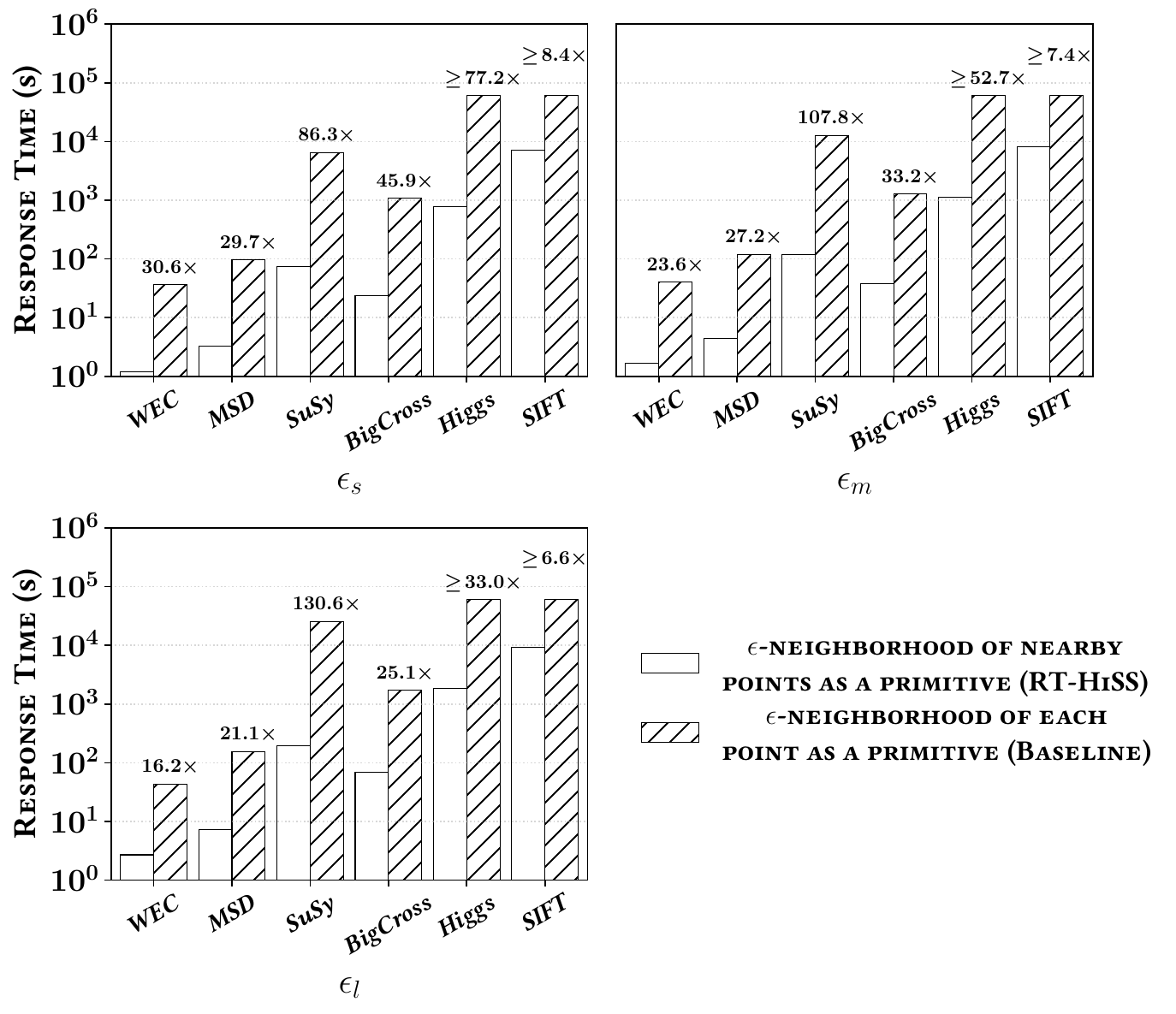}

    \caption{Response time (s) on a logarithmic scale of \ouralg when a primitive is defined by the combined $\epsilon$-neighborhood of nearby points compared to that of each point as a primitive.}
   \label{fig:onePointPerPrim}
\end{figure}

All \dss algorithms using RT cores represent the $\epsilon$-neighborhood  of each point in the dataset as a primitive in the \optix scene~\cite{zhu2022rtnn, nagarajan2023rt}. The number of primitives in the scene directly affects the number of ray–primitive intersections and consequently, the overall performance of the algorithm. In contrast, we identify nearby points in the data space and represent the $\epsilon$-neighborhood  of all the nearby points as a single primitive (Section~\ref{sec:coarseIndex}).

Figure~\ref{fig:onePointPerPrim} shows the performance of \ouralg on a logarithmic scale when representing the $\epsilon$-neighborhood  of each point as a primitive compared to our proposed approach of grouping the $\epsilon$-neighborhood of nearby points together into a single bounding sphere. Across all datasets and search distances, performance improves significantly with this optimization. We observe a minimum and maximum speedup of $16.2\times$ and $130.6\times$ for \wave and \susy with $\epsilon_l$, respectively.

These results indicate that grouping nearby points together has a more pronounced impact on computationally expensive datasets (i.e., \higgs). This is because such datasets typically incur a high number of ray–primitive intersections when the $\epsilon$-neighborhood  of each point is represented as a primitive, which directly increases the workload on RT cores during the indexing phase of the two-pass ray tracing approach. By identifying nearby points, and representing the $\epsilon$-neighborhoods  of nearby points as a single primitive, we substantially reduce the number of ray-primitive intersections, thereby reducing the cost of indexing. This yields an average speedup of 42.4$\times$ across all datasets and search distances.

\noindent\textbf{Takeaway \#1: }\emph{Grouping points together is a primary factor in enabling the effective use of RT cores for indexing large, high-dimensional datasets. Overall, representing the $\epsilon$-neighborhood of nearby points as a single bounding sphere significantly improves overall performance.}

\subsection{Ablation Study \#1: Point Refinement Kernel}\label{sec:ablationPoint}
\begin{table}[t]
\setlength{\tabcolsep}{0.5em}

    \caption{Experiment configurations used in our ablation studies. Here, \shq, \shp, and \shno are point comparison variants whereas uncompressed result mask copy and key-value copy are result representation variants.}
    \centering
    \footnotesize

    \begin{tabularx}{\columnwidth}{l Y Y Y Y Y}
        \hline
        &\multicolumn{2}{c}{Point Comparison}&\multicolumn{3}{c}{Result \& Copying Variants}\\
        % \cline{2-6}
         Experimental Config. & Query Shared Mem. & Primitive Shared Mem. &Compr- essed  & Uncompr-essed & Key-Value \\
         \hline
            \ouralg & \cmark & \cmark & \cmark & \xmark & \xmark \\
            \hline
            \multicolumn{6}{c}{{Ablation Study \#1: Point Comparison (Section~\ref{sec:ablationPoint})}} \\
            \hline
             \shq & \cmark & \xmark & \cmark & \xmark & \xmark \\
             \shp & \xmark & \cmark & \cmark & \xmark & \xmark \\
             \shno & \xmark & \xmark & \cmark & \xmark & \xmark \\
            \hline
            \multicolumn{6}{c}{{Ablation Study \#2: Result Representation (Section~\ref{sec:ablationResultRep})}} \\
            \hline
         Uncomp. mask  & \cmark & \cmark & \xmark & \cmark & \xmark \\
           Key-value copy & \cmark & \cmark & \xmark & \xmark & \cmark \\            
         \hline
    \end{tabularx}
    \label{tab:expConfig}
\end{table}

\begin{figure}[!t]
\centering

\includegraphics[width=\columnwidth]{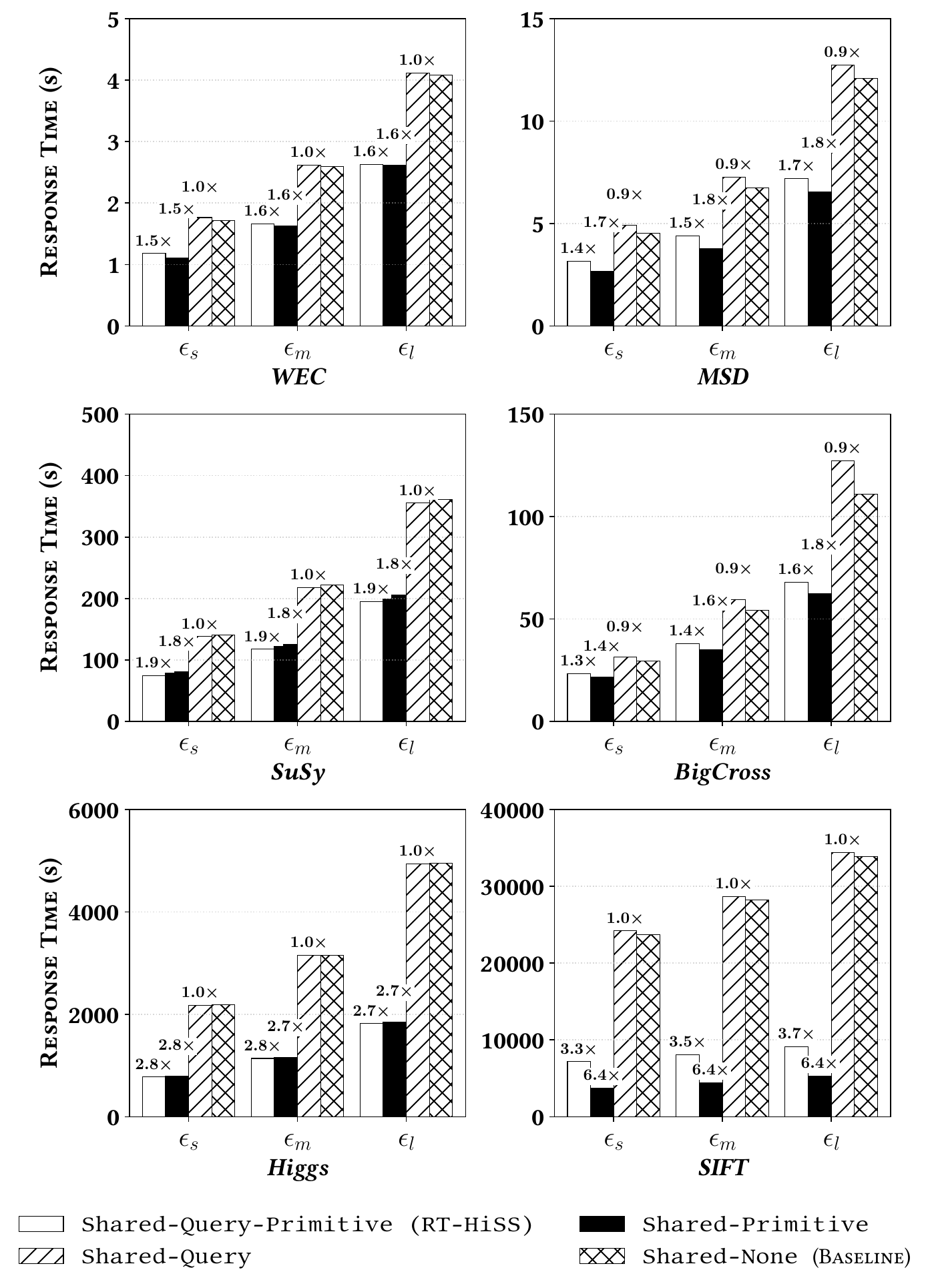}
 % \revanth{Update figure.}

    \caption{Response time (s) of \ouralg and the point comparison variants outlined in Section~\ref{sec:ablationPoint}. Here, \shno is used as the baseline, and the speedup of each variant relative to the baseline is shown.}
    % \mike{Increase size 3 rows x 2 cols}
   \label{fig:pointCompVariant}
\end{figure}

We conduct an ablation study to examine the impact of reading query and primitive points from shared memory, by disabling optimizations in \ouralg (see Table~\ref{tab:expConfig}). Figure~\ref{fig:pointCompVariant} shows the ablation study results where we compare to the baseline, \shno, in which both query and primitive points are loaded from global memory. Reading only query points from shared memory (\shq) yields a slowdown over the baseline on some datasets (i.e., \wave, \msd, \bigcross, and \sift) due to insufficient data reuse and synchronization overhead.  Reading only primitive points, \shp, outperforms \shno in all experimental scenarios yielding an average speedup of 2.64$\times$. Similarly, \shqp (\ouralg) outperforms \shno across all datasets and search distances with an average speedup of 2.12$\times$. 

In summary, loading primitive and query points into shared memory improves performance by reducing off-chip global memory accesses. However, we show that shared memory should be used judiciously to avoid performance degradation.

\subsection{Ablation Study \#2: Result Representation}\label{sec:ablationResultRep}

We examine the impact of different result representations on the number of batches and the response time of \ouralg.  As described in Table~\ref{tab:expConfig}, we disable optimizations in \ouralg which uses a compressed result mask. We compare this to two variants: $(i)$ not compressing the result mask; and $(ii)$ using key-value pairs to represent the result set.

\subsubsection{{\bf Impact on the Number of Batches}}\label{sec:batching}

\begin{table}[t]
\caption{The number of batches required for result representation variants outlined in Section~\ref{sec:ablationResultRep} with $\epsilon_l$. The ratio of the number of batches for the key-value copy compared to \ouralg is shown in parentheses. Fewer batches are better.}
\label{tab:unCompressedCopy}
\centering
\setlength{\tabcolsep}{2pt}
\footnotesize
% \revanth{Show results only for $\epsilon_l$}
 % \revanth{Update table.}
\begin{tabularx}{\columnwidth}{l l R R R}
\hline
Dataset & $\epsilon$ &  Comp. Result Mask Copy (\ouralg) & Uncomp. Result Mask Copy & Key-value Copy \\
\hline
\wave & $\epsilon_l$ & 1  & 1  & 13 (13.00 $\times$)\\
% \hline
\msd &$\epsilon_l$ & 1  & 1  & 47 (47.00$\times$)\\
% \hline
\susy & $\epsilon_l$ & 100  & 100 & 5,563 (55.63$\times$)\\
% \hline
\bigcross & $\epsilon_l$  & 16  & 16 & 860 (53.75$\times$)\\
% \hline
\higgs & $\epsilon_l$ & 896  & 896 & 50,791 (56.69$\times$)\\

\sift & $\epsilon_l$  & 1,638 & 1,638& 91,621 (55.93$\times$) \\

\hline

\end{tabularx}
\end{table}

Table~\ref{tab:unCompressedCopy} compares the number of batches required to process all queries for different result representation strategies using the $\epsilon_l$ search distance, where we observe that using a result mask significantly reduces the number of batches compared to key-value copies. The compressed and uncompressed result mask variants use the same number of batches, and require 1--1,638 batches, whereas the key-value copy requires 13 to $\gtrsim$90K batches, with \wave and \sift requiring the minimum and maximum number of batches, respectively. 

The key–value copy strategy requires significantly more batches, as each result element uses two 32-bit integers, whereas \ouralg uses only one bit per point comparison, achieving a 64$\times$ compression ratio (Section~\ref{sec:copyResults}). This difference enables \ouralg to execute more queries per batch and enables GPU resource saturation.

\subsubsection{{\bf Impact of Batching on Response Time}}\label{sec:resultCopyTimeEndToEnd}

\begin{figure}[!t]
\centering
 % \revanth{Update figure.}
\includegraphics[width=\columnwidth]{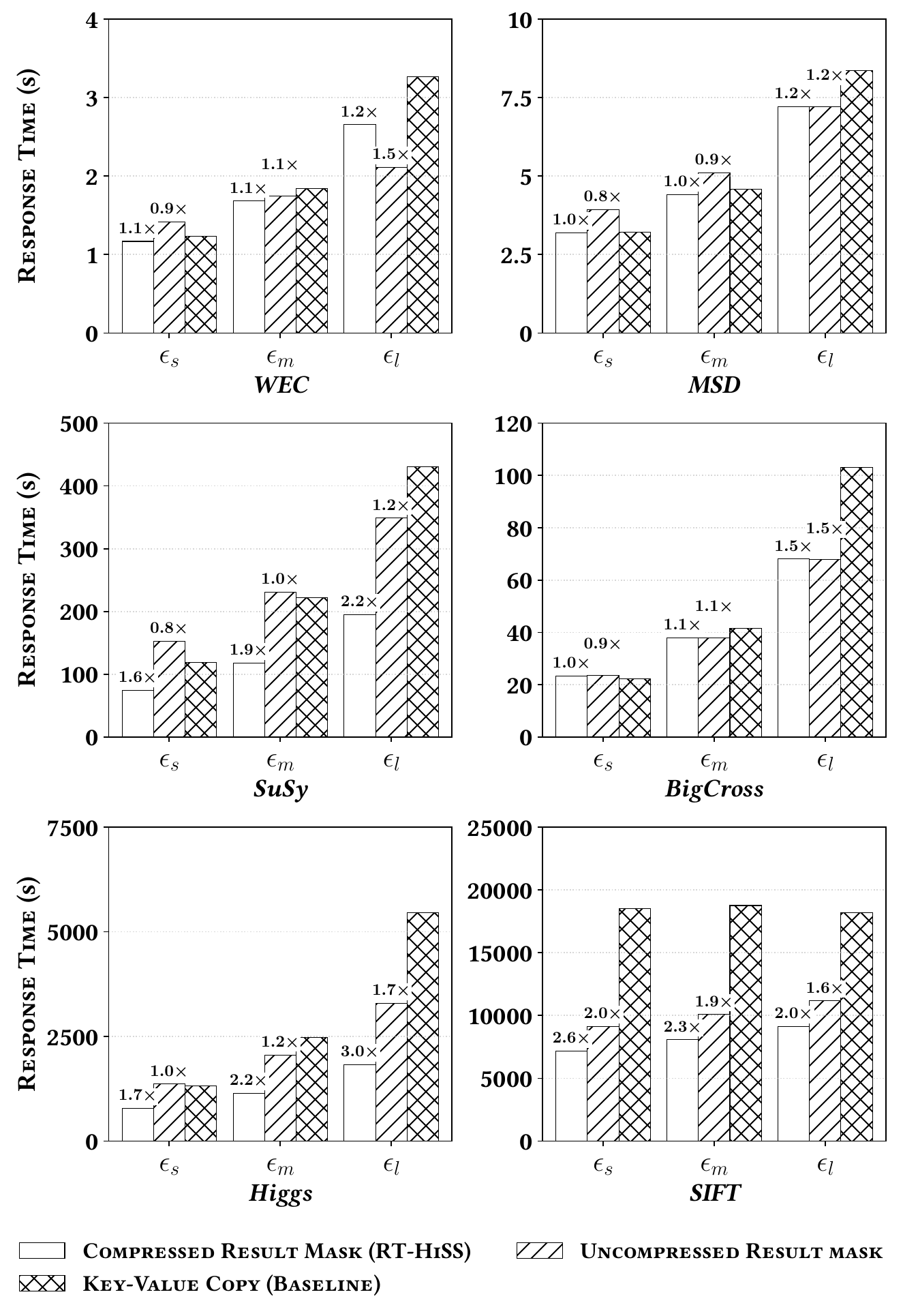}
    \caption{Response time (s) of \ouralg compared to the result representation variants  (Section~\ref{sec:ablationResultRep}). The speedup of \ouralg and the uncompressed result mask copy relative to the baseline (key-value pairs) is shown.}
    % \mike{Increase size}

   \label{fig:copyVariant}
\end{figure}

 The overall performance of the algorithm is directly affected by the number of batches as each batch requires two RT kernel executions (Section~\ref{sec:twoPassApproach}). Figure~\ref{fig:copyVariant} shows the response time of \ouralg compared to those in Table~\ref{tab:expConfig}. We observe that \ouralg outperforms key-value copy across all evaluation scenarios with an average speedup of 1.65$\times$. Similarly, except for a few evaluation scenarios, the compressed result mask copy (\ouralg) outperforms the uncompressed result mask copy method across all datasets and search distances with an average speedup of 1.24$\times$. This improvement is due to reduced data transfer time resulting from result mask compression.

\noindent\textbf{Takeaway \#2: }\emph{While the representation of the result set may seem inconsequential to performance, the larger space requirements of storing key-value pairs necessitates additional batches to process all of the queries which directly corresponds to the additional overhead induced by the two-pass ray tracing approach during batching. In addition to using a result mask, compression reduces the result set size and improves data transfer time, which further improves performance.}

\subsection{Comparison with the State-of-the-art}
% \revanth{Maybe divide this into subsubsections. One for Cuda-based and another for RT-based.}

\begin{table*}[!t]

\caption{Response time (s) of \ouralg compared to the SOTA algorithms outlined in Section~\ref{sec:refAlg}, presented in the order of increasing mean speedup over the SOTA, with speedups shown in parentheses. The best time is shown in bold.}
\label{tab:expStateofTheArt}
\centering
\setlength{\tabcolsep}{2pt}
\footnotesize
% \revanth{Update table. Fix order of datasets. Sort by number of points or dimensionality.}
\begin{tabularx}{\textwidth}{L l r r r r r r}
\hline
Dataset & $\epsilon$ & \ouralg & \mikealg~\cite{gdsJoin} & \brianalg~\cite{coss} & \makecell[r]{\cuvsalg \\(NVIDIA)~\cite{cuvs}}  & \makecell[r]{\pyTorchAlg \\ (Facebook)~\cite{pytorch3d}}  & \GTS~\cite{GTS} \\
\hline

\wave & $\epsilon_s$ & \textbf{1.16} & 1.51 (1.31$\times$)  & 1.93 (1.67$\times$)          & 14.88 (12.87$\times$) & 74.99 (64.87$\times$)       & 3,739.45 (3,234.81$\times$)\\
\wave & $\epsilon_m$ & \textbf{1.67} & 3.27 (1.95$\times$)  & 2.92 (1.74$\times$)          & 14.88 (8.89$\times$)  & 133.43 (79.71$\times$)      & 5,006.74 (2,990.88$\times$)\\
\wave & $\epsilon_l$ & 2.71          & 7.10 (2.62$\times$)  & \textbf{1.82 (0.67$\times$)} & 14.88 (5.50$\times$)  & 5,790.01 (2,139.69$\times$) & 5,908.50 (2,183.48$\times$)\\
\hline

\msd & $\epsilon_s$ & \textbf{3.18} & 5.48 (1.72$\times$)  & 5.00 (1.57$\times$)  & 49.20 (15.46$\times$) & 544.75 (171.20$\times$)      & 14,404.56 (4,526.89$\times$)\\
\msd & $\epsilon_m$ & \textbf{4.41} & 9.66 (2.19$\times$)  & 7.38 (1.67$\times$)  & 49.20 (11.17$\times$) & 1,378.92 (313.04$\times$)    & 19,121.26 (4,340.81$\times)$\\
\msd & $\epsilon_l$ & \textbf{7.24} & 19.80 (2.74$\times$) & 15.75 (2.18$\times$) & 49.20 (6.80$\times$)  & 30,243.18 (4,180.12$\times$) & 25,743.88 (3,558.24$\times$)\\
\hline

\susy & $\epsilon_s$ & 74.48           & \textbf{52.14 (0.70$\times$)} & 94.63 (1.27$\times$)  & 9,616.54 (129.11$\times$) & 2,668.60 (35.83$\times$)       & 31,704.42 (425.65 $\times$) \\
\susy & $\epsilon_m$ & \textbf{117.31} & 122.55 (1.04$\times$)         & 196.50 (1.68$\times$) & 9,616.54 (81.97$\times$)  & 14,846.70 (126.56$\times$)     & 50,823.54 (433.24$\times$)\\
\susy & $\epsilon_l$ & \textbf{195.39} & 424.32 (2.17$\times$)         & 488.61 (2.50$\times$) & 9,616.54 (49.22$\times$)  & \timeout ($\ge$307.08$\times$) & \timeout ($\ge$307.08$\times$) \\
\hline

\bigcross & $\epsilon_s$ & \textbf{23.20} & 37.55 (1.62$\times$)  & 94.63 (4.08$\times$)  & 54,943.70 (2,368.26$\times$) & \timeout ($\ge$2,586.21$\times$)  & 37,585.78 (1,620.08$\times$)\\
\bigcross & $\epsilon_m$ & \textbf{37.87} & 143.88 (3.80$\times$) & 196.50 (5.19$\times$) & 54,943.70 (1,451.04$\times$) & \timeout ($\ge$1,584.58$\times$) & 51,911.63 (1,370.97$\times$)\\
\bigcross & $\epsilon_l$ & \textbf{67.90} & 623.97 (9.19$\times$) & 488.61 (7.20$\times$) & 54,943.70 (809.21$\times$)   & \timeout ($\ge$883.68$\times$)   &\timeout ($\ge$883.68$\times$)\\
\hline

\higgs & $\epsilon_s$  & \textbf{775.55}   & 1,801.58 (2.32$\times$) & 3,362.11 (4.34$\times$) & 46,551.90 (60.02$\times$) & 26,526.54 (34.20$\times$)     &\timeout ($\ge$77.36$\times$) \\
\higgs & $\epsilon_m$  & \textbf{1,139.14} & 2,101.15 (1.84$\times$) & 3,292.26 (2.89$\times$) & 46,551.90 (40.87$\times$) & \timeout ($\ge$52.67$\times$) &\timeout ($\ge$52.67$\times$) \\
\higgs & $\epsilon_l$  & \textbf{1,818.49}  & 2,819.33 (1.55$\times$) & 4,460.49 (2.45$\times$) & 46,551.90 (25.60$\times$) & \timeout ($\ge$32.99$\times$) &\timeout ($\ge$32.99$\times$)\\
\hline

\sift & $\epsilon_s$  & \textbf{7,169.15} & 7,732.33 (1.08$\times$)  & \timeout ($\ge$8.37$\times$) & 43,250.30 (6.03$\times$) & \timeout ($\ge$8.37$\times$) & \timeout ($\ge$8.37$\times$) \\ 
\sift & $\epsilon_m$  & \textbf{8,070.70} & 9,232.53 (1.14$\times$)  & \timeout ($\ge$7.43$\times$) & 43,250.30 (5.36$\times$) & \timeout ($\ge$7.43$\times$) & \timeout ($\ge$7.43$\times$) \\
\sift & $\epsilon_l$  & \textbf{9,134.09} & 11,290.50 (1.24$\times$) & \timeout ($\ge$6.57$\times$) & 43,250.30 (4.74$\times$) & \timeout ($\ge$6.57$\times$) & \timeout ($\ge$6.57$\times$) \\
\hline

\multicolumn{2}{l}{\textbf{Speedup Range}} 
& \textbf{[1.00, 1.00]$\times$} 
& \textbf{[0.70, 9.19]$\times$} 
& \textbf{[0.67, 8.37]$\times$} 
& \textbf{[4.74, 2,368.26]$\times$} 
& \textbf{[34.20, 4,180.12]$\times$}  
& \textbf{[425.65, 4,526.89]$\times$} \\

\hline
\end{tabularx}
\end{table*}

We compare \ouralg with other GPU SOTA algorithms. Table~\ref{tab:expStateofTheArt} reports the performance of \ouralg against the reference algorithms described in Section~\ref{sec:refAlg}.

\noindent\textbf{Indexing Approaches:} We compare to the SOTA GPU algorithms that employ an index to prune the search space. We observe that \ouralg outperforms \mikealg achieving a speedup up to 9.19$\times$ across all datasets (\ouralg only yielded a slowdown for \susy dataset with $\epsilon_s$ search distance). Similarly, \ouralg outperforms \brianalg across all datasets and $\epsilon$ distances, except for the \wave dataset with the $\epsilon_l$ search distance, yielding a maximum speedup of 8.37$\times$.

\ouralg also outperforms \pyTorchAlg across all datasets and search distances achieving speedup spanning 34.20$\times$ to 4,180.12$\times$ across all evaluation scenarios. Similarly, \ouralg outperforms \GTS across all datasets and search distances. \GTS reaches the timeout limit on several experiments, and is not competitive with \ouralg.

\noindent\textbf{Brute-Force Approaches:} Lastly, we compare to \cuvsalg which is brute-force approach by NVIDIA~\cite{cuvs}. \ouralg outperforms \cuvsalg across all datasets and search distances, with minimum and maximum speedup of 4.74$\times$ and 2,368.26$\times$, respectively.

% Overall, these results show that \ouralg achieves a minimum speedup of 0.79$\times$ compared to \mikealg and a maximum speedup of at least 5K$\times$ compared to \GTS. 

Overall, we find that \ouralg largely yields significant speedups as it addresses majority of the challenges of processing data-dependent workloads and scales well with dataset size. \mikealg uses a grid-based indexing method for pruning the $\epsilon$-neighborhood  and employs an efficient batching scheme with overlapping host and GPU computation to achieve high performance. \brianalg uses metric-based indexing along with an efficient batching scheme. \cuvsalg is an optimized brute-force algorithm that does not use a result estimator and allocates $|D|^2$ elements to store the result set, which limits the number of queries processed per batch and introduces batching overhead. \pyTorchAlg uses a maximum neighbors per query parameter to allocate memory for results, where the parameter value is set based on the selectivity level for each $\epsilon$ distance. \GTS employs a pivot-based tree structure and performs poorly for large datasets and larger search distances.

% \revanth{Two-pass RT core approach enables a near-perfect load balancing scheme?}
\noindent\textbf{Takeaway \#3: }\emph{The two-pass RT core approach yields two major advantages: $(i)$ it enables fast indexing and searching; and $(ii)$ it enables an efficient batching scheme as it bounds the expected number of results based on the query-primitive intersections, which in turn results in saturating GPU resources. This enables significant speedups over the SOTA.}

\subsection{Comparison with RT-based State-Of-the-art ($d=3$)}

\begin{table}[t]
\caption{Response time (s) of \ouralg compared to \textsc{rtnn} for $d=3$ datasets. The speedup of \ouralg over \rtnn is listed in the speedup column. The best time is in bold.}
\label{tab:rtnnComp}
\centering
\setlength{\tabcolsep}{2pt}
\footnotesize
 
\begin{tabularx}{\columnwidth}{X r r r r r r}
\hline
Dataset  & $d$ & $|D|$ & $\epsilon_s$ &  \ouralg & \rtnn & Speedup \\
\hline
\iono~\cite{gowanlock2016exploiting} & 3 &5,159,737 & 0.457142 & \textbf{6.22} & 8.79 & 1.41$\times$\\

\gaia~\cite{gaia2018gaia} & 3 & 5,000,000 & 0.374163 & \textbf{5.22} & 8.57 & 1.64$\times$\\

\uniform & 3 & 5,000,000 & 0.023206 & \textbf{4.58} & 8.62 & 1.88$\times$\\

\expo & 3 & 5,000,000 & 0.001170 & \textbf{7.10} & 8.54 & 1.20$\times$\\
\hline

\textbf{Speedup Range } & & & & & & \textbf{[1.20, 1.88]$\times$} \\
\hline

\end{tabularx}
\end{table}

 % This investigation allows us to understand the impact of our proposed optimizations on low-dimensional ($d=3$) datasets.

We compare \ouralg to \rtnn to examine the performance on low dimensional datasets. Table~\ref{tab:rtnnComp} lists two real-world (\iono and \gaia) and two synthetic (\uniform and \expo) $d=3$ datasets used to compare the performance of \ouralg relative to \rtnn using the $\epsilon_s$ search distance.\footnote{We do not compare with $\epsilon_m$ and $\epsilon_l$ as \rtnn does not have a batching scheme and runs out of global memory on our platform.} The synthetic exponential dataset is generated using a rate parameter $\lambda=40$. We observe that \ouralg outperforms \rtnn across all datasets with speedups up to 1.88$\times$.

\ouralg is designed to accelerate high-dimensional distance similarity searches using RT cores; however, there are two design decisions for high-dimensional searches that reduce the performance of \ouralg in the low-dimensional setting as described above.
$(i)$ Recall that \ouralg first identifies candidate neighbor points during the RT indexing step and later refines in a separate CUDA refinement kernel (Section~\ref{sec:twoPassApproach}). The performance of \ouralg for low dimensional \dss could be improved by refining within the RT kernel itself instead of using a separate kernel. $(ii)$ Similarly, the result set compression optimization benefits high-dimensional \dss but introduces overhead for low-dimensional \dss. 

\subsection{\ouralg Indexing Efficiency}

Euclidean distance calculations are often the most expensive step in similarity searches due to their quadratic time complexity~\cite{donnelly2024multi}. We compare the reduction in the number of Euclidean distance calculations performed by \ouralg and \mikealg relative to \cuvsalg using the $\epsilon_l$ search distance to understand the efficiency of the employed pruning method in mitigating these distance calculations. As shown in Table~\ref{tab:pointComparisons}, \ouralg indexing using the BVH tree in 3 dimensions performs fewer distance calculations relative to \mikealg using a grid indexing in 6 dimensions, except for the \susy dataset where \ouralg is slower than \mikealg (Table~\ref{tab:expStateofTheArt}). This shows that \ouralg prunes the search space more efficiently despite indexing in only 3 dimensions. We observe a high degree of pruning for \wave, \msd, \susy, and \bigcross ($\geq$ 64.7\%), whereas low pruning is observed on \higgs (37.9\%). Even though no pruning occurs on \sift, \ouralg and \mikealg outperforms \cuvsalg across all search distances due to locality aware computation and aborting the search early, among other optimizations.

\begin{table}[!t]
 % \revanth{Update table.}
% \caption{The number of Euclidean distance calculations performed by \ouralg, \mikealg, and \cuvsalg using $\epsilon_l$ search distance. The reduction in distance calculations due to the indexing is shown in parenthesis. Lower number of distance calculations are better.}

\caption{The reduction in the number of Euclidean distance calculations due to the employed pruning method for \ouralg and \mikealg relative to \cuvsalg. Higher is better.}

\label{tab:pointComparisons}
\centering
\setlength{\tabcolsep}{2pt}
\footnotesize
\begin{tabularx}{\columnwidth}{X R R}
\hline
Dataset & \ouralg & \mikealg \\
\hline
\wave & \textbf{76.4\%} & 74.4\%  \\
\msd & \textbf{71.7\%} & 67.9\%   \\
\susy & 64.7\% & \textbf{71.2\%}  \\
\bigcross & \textbf{99.2\%} & 99.0\%  \\
\higgs & \textbf{37.9\%} & 0.8\%  \\
\sift & \textbf{0.0\%} & \textbf{0.0\%}  \\
\hline
\end{tabularx}
\end{table}

\subsection{\ouralg Profiler Statistics}\label{sec:profilerResults}

\begin{table}[!t]
 % \revanth{Update table.}
\caption{Profiler statistics for the refinement kernel comparing \ouralg to \shno (baseline). The increase in compute throughput and the reduction in global instructions are shown in the ratio and reduction columns, respectively.}

\label{tab:profiler}
\centering
\setlength{\tabcolsep}{3pt}
\footnotesize
% \begin{tabularx}{\columnwidth}{|l|R|R|R|R|R|R|}
\begin{tabularx}{\columnwidth}{l| R R R |R R R}
\hline
Dataset
& \multicolumn{3}{c|}{Compute Throughput}
& \multicolumn{3}{c}{Global Memory Instructions}\\
\cline{2-7}

& \ouralg & \shno & Ratio
& \ouralg & \shno & Reduction \\

\hline
\wave      & 72.75\%  & 28.97\% & 2.51$\times$ & 1.18 G & 27.16 G & 95.66\%  \\
\msd       & 61.00\%  & 26.45\% & 2.31$\times$ & 7.22 G & 92.76 G & 92.22\%  \\
\susy      & 77.81\%  & 37.82\% & 2.06$\times$ & 0.72 G & 37.43 G & 98.08\%  \\
\bigcross  & 60.23\%  & 23.66\% & 2.55$\times$ & 3.41 G & 43.98 G & 92.25\%  \\
\higgs     & 84.22\%  & 27.90\% & 3.02$\times$ & 0.93 G & 56.38 G & 98.35\%  \\
\sift      & 53.27\% & 12.28\% & 4.34$\times$ & 12.85 G & 89.49 G & 85.64\% \\
\hline
\end{tabularx}
\end{table}

% Old: Avg of shared memory load request : 8.05\%

We profile the GPU kernel that refines candidate points to examine the impact of our proposed optimizations using Nvidia Nsight Compute. At present, RT kernels cannot be profiled using NVIDIA Nsight Compute. We compare \ouralg to the baseline variant that does not use shared memory tiling (\shno). Table~\ref{tab:profiler} compares the compute throughput and the number of global memory instructions for all datasets with $\epsilon_l$.

 We observe that compute throughput increases significantly when shared memory tiling is employed to read point data. Specifically, \ouralg achieves 2.06$\times$ to 4.34$\times$ compute throughput improvement compared to the baseline. Next, examining the number of global memory loads, we observe a reduction of 85.64\% to 98.35\% in \ouralg compared to the baseline. This highlights the benefit of staging data in shared memory and minimizing off-chip global memory access requests. Lastly, we find that the number of shared memory load bank conflicts spans 5.13\% to 42.89\% (average of 13.86\%) across all datasets. These results demonstrate that the data path is well-optimized through efficient shared memory tiling and coalesced memory accesses (Section~\ref{sec:reorderPoints}).

\subsection{Time Distribution Analysis}\label{sec:timeDistrib}

\begin{figure}[!t]
\centering
\includegraphics[width=\columnwidth]{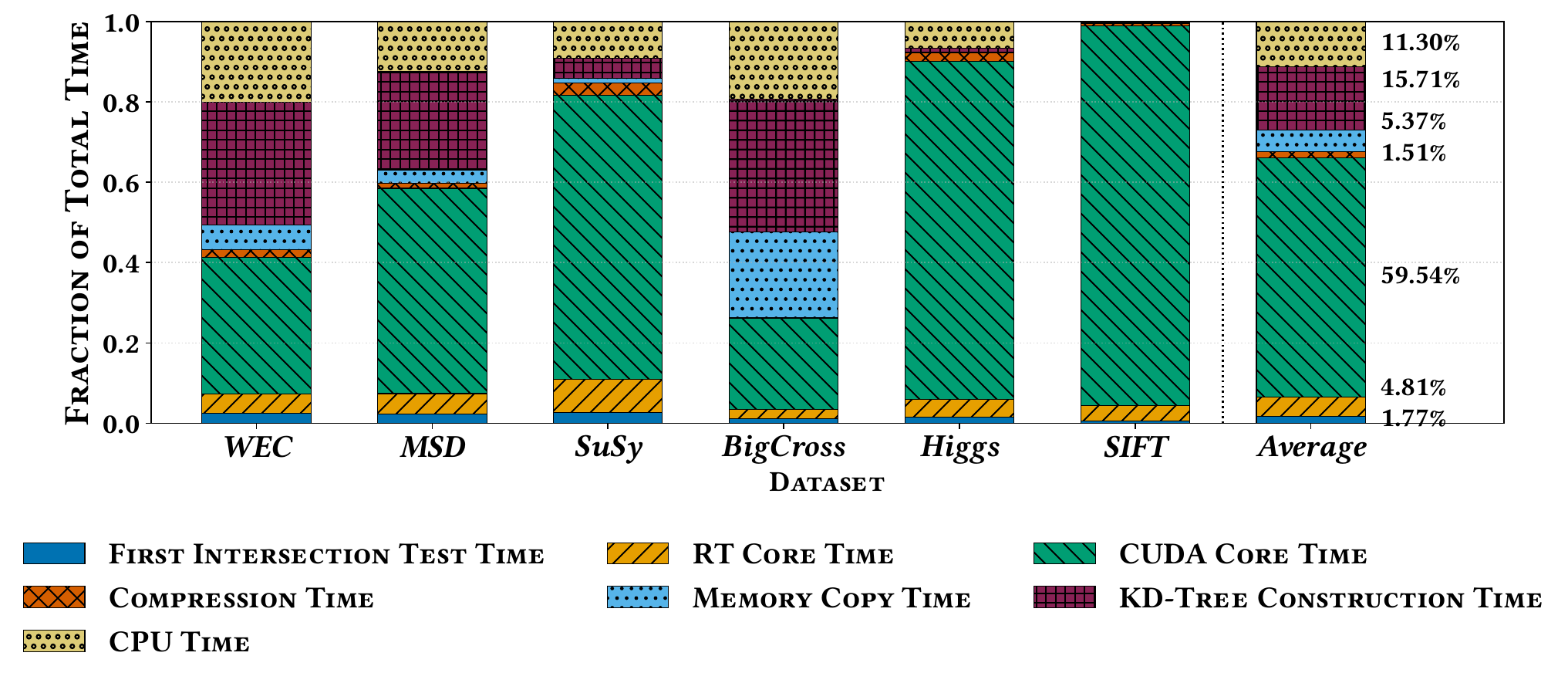}
 % \revanth{Update figure.}
    \caption{Normalized response time of \ouralg components, averaged across all search distances. The ``Average’’ bar represents the average time spent in each component across all datasets and search distances.}
   \label{fig:timeDistrib}
\end{figure}

We examine the time distribution of each component of \ouralg. Figure~\ref{fig:timeDistrib} illustrates the fraction of total time spent in each component. The total time is averaged across all search distances ($\epsilon_s$, $\epsilon_m$, and $\epsilon_l$) and normalized to the range [0,1].

We observe that across all datasets, CUDA core time accounts for the majority of the total execution time (average of 59.54\%) followed by the kd-tree construction time (average of 15.71\%). CPU time (average of 11.30\%) includes the time to identify bounding spheres of nearby points, and other host-side tasks. The FIT and two-pass RT-core approach together account for only an average of 6.58\% of the total time, which is minimal compared to the other components. Thus, RT cores provide a lightweight method for efficient indexing and searching, while also enabling a batching scheme with near-perfect load balancing across batches.

\section{Discussion \& Conclusion}\label{sec:conclusion}

To address diverse workloads, HPC systems contain numerous processor types including tensor cores, ray tracing cores, data processing units, FPGAs, among others. We examine the design space for high-dimensional distance similarity searches and find that RT cores can be used for hardware-accelerated index searches. We address the inherent hardware limitations of RT cores with an algorithmic approach that leverages RT cores for indexing followed by CUDA cores for refinement for high-dimensional \dss, yielding significant performance gains over state-of-the-art GPU algorithms. 

The paper focuses on accelerating distance similarity searches on a single-GPU system; however, our algorithm (\ouralg) and the proposed optimizations can be directly translated into multi-GPU systems and HPC environments with multiple nodes. In particular, after the first intersection test, \ouralg identifies the number of distance calculations required for each query based on the number of ray-bounding sphere intersections. This is then used to support the proposed batching scheme. In multi-GPU or HPC environments, instead of processing one batch at a time, work could be asynchronously offloaded to other GPUs or nodes while employing all the optimizations discussed in the paper. This improves scalability for large datasets and can also be exploited for near-real time distance similarity searches.

Our future work includes other types of similarity searches. Similarly to distance similarity searches, the high dimensional $k$-nearest neighbors  problem is also constrained by the inherent limitation of RT cores to three dimensions. This can be addressed by extending \ouralg to translate the $k$-nearest neighbors problem into a recursive similarity search problem that expands the initial search radius ($\epsilon$) until $k$ neighbors are found. This approach yields the exact $k$-nearest neighbors while indexing in three dimensions. All the optimizations proposed in the paper can be applied to this problem as well. 

\revanth{Changed last line.}
\ouralg is designed for large high-dimensional datasets and using it for small or low-dimensional datasets with insufficient work leads to resource underutilization including other limitations discussed in the paper, thereby resulting in degraded performance. We will examine this limitation in addition to extending \ouralg to support other distance similarity searches in future work.

% Dr.Gowanlock to update this section for final submission.
% MG: I added the grant to the spot where it goes in the title.
\section*{Acknowledgment}
% This material is based upon work supported by the National Science Foundation under Grant No. 2042155. 
The Chameleon testbed supported by the National Science Foundation~\cite{keahey2020lessons} was used as part of our Artifact Description and Evaluation appendices.

\bibliographystyle{IEEEtran}
\bibliography{bibliography}

@article{datasetsusyhiggs,
  title={{Searching for Exotic Particles in High-Energy Physics with Deep Learning}},
  author={Baldi, Pierre and Sadowski, Peter and Whiteson, Daniel},
  journal={Nature Communications},
  volume={5},
  pages={4308},
  year={2014},
  publisher={Nature Publishing Group},
  url = {https://doi.org/10.1038/ncomms5308}
}

@article{ackermann2012streamkm,
  title={{{StreamKM++} A Clustering Algorithm for Data Streams}},
  author={Ackermann, Marcel R and M{\"a}rtens, Marcus and Raupach, Christoph and Swierkot, Kamil and Lammersen, Christiane and Sohler, Christian},
  journal={Journal of Experimental Algorithmics (JEA)},
  volume={17},
  pages={2--1},
  year={2012},
  publisher={ACM New York, NY, USA},
  url = {https://doi.org/10.1145/2133803.2184450}
}

@inproceedings{datasetmsd,
author = {Bertin-Mahieux, Thierry and Ellis, Daniel and Whitman, Brian and Lamere, Paul},
year = {2011},
month = {01},
pages = {591-596},
title = {{The Million Song Dataset.}},
  booktitle = {{Proceedings of the 12th International Conference on Music Information
	Retrieval ({ISMIR} 2011)}},
url = {https://doi.org/10.7916/D8NZ8J07}
}

@misc{datasetSift,
  author       = {Fu, Xiping and McCane, Brendan and Mills, Steven and Albert, Michael and Szymanski, Lech},
  title        = {{SIFT10M}},
  year         = {2016},
  howpublished = {UCI Machine Learning Repository},
  note         = {{DOI}: https://doi.org/10.24432/C5S603}
}

@inproceedings{datasetwave,
  title={{A Detailed Comparison of Meta-Heuristic Methods for Optimising Wave Energy Converter Placements}},
  author={Neshat, Mehdi and Alexander, Bradley and Wagner, Markus and Xia, Yuanzhong},
  booktitle={Proceedings of the Genetic and Evolutionary Computation Conference},
  pages={1318--1325},
  year={2018},
  url = {https://doi.org/10.1145/3205455.3205492}
}

@inproceedings{gdsJoin,
  title={{GPU-Accelerated Similarity Self-Join For Multi-Dimensional Data}},
  author={Gowanlock, Michael and Karsin, Ben},
  booktitle={Proceedings of the 15th International Workshop on Data Management on New Hardware},
  pages={1--9},
  year={2019},
  url = {https://doi.org/10.1145/3329785.3329920}
}

@inproceedings{coss,
  title={{A Coordinate-Oblivious Index for High-Dimensional Distance Similarity Searches on the {GPU}}},
  author={Donnelly, Brian and Gowanlock, Michael},
  booktitle={Proceedings of the 34th ACM International Conference on Supercomputing},
  pages={1--12},
  year={2020},
  url = {https://doi.org/10.1145/3392717.3392768}
}

@misc{cuvs,
  title        = {{cuVS: GPU-Accelerated Vector Search and Clustering Library}},
  author       = {NVIDIA},
  howpublished = {\url{https://github.com/rapidsai/cuvs}},
  year         = {2025},
  note         = {Version 25.12.00, Apache-2.0 License}
}

@article{GTS,
  title={{{GTS}: GPU-based Tree Index for Fast Similarity Search}},
  author={Zhu, Yifan and Ma, Ruiyao and Zheng, Baihua and Ke, Xiangyu and Chen, Lu and Gao, Yunjun},
  journal={Proceedings of the ACM on Management of Data},
  volume={2},
  number={3},
  pages={1--27},
  year={2024},
  publisher={ACM New York, NY, USA},
  url = {https://doi.org/10.1145/3654945}
}

@inproceedings{guttman1984r,
  title={{R-Trees: A Dynamic Index Structure for Spatial Searching}},
  author={Guttman, Antonin},
  booktitle={Proceedings of the 1984 ACM SIGMOD International Conference on Management of Data},
  pages={47--57},
  year={1984},
  url = {https://doi.org/10.1145/602259.602266}
}

@article{jackins1980oct,
  title={{Oct-Trees and Their Use in Representing Three-Dimensional Objects}},
  author={Jackins, Chris L and Tanimoto, Steven L},
  journal={Computer Graphics and Image Processing},
  volume={14},
  number={3},
  pages={249--270},
  year={1980},
  publisher={Elsevier},
  url = {https://doi.org/10.1016/0146-664X(80)90055-6}
}

@article{dolatshah2015ball,
  title={{Ball*-tree: Efficient Spatial Indexing for Constrained Nearest-Neighbor Search in Metric Spaces}},
  author={Dolatshah, Mohamad and Hadian, Ali and Minaei-Bidgoli, Behrouz},
  journal={arXiv preprint arXiv:1511.00628},
  year={2015},
  url = {https://doi.org/10.48550/arXiv.1511.00628}
}

@inproceedings{zhu2022rtnn,
  title={{{RTNN}: Accelerating Neighbor Search Using Hardware Ray Tracing}},
  author={Zhu, Yuhao},
  booktitle={Proceedings of the 27th ACM SIGPLAN Symposium on Principles and Practice of Parallel Programming},
  pages={76--89},
  year={2022},
  url = {https://doi.org/10.1145/3503221.3508409}
}

@inproceedings{nagarajan2023rt,
  title={{{RT-kNNS} unbound: Using {RT} Cores to Accelerate Unrestricted Neighbor Search}},
  author={Nagarajan, Vani and Mandarapu, Durga and Kulkarni, Milind},
  booktitle={Proceedings of the 37th International Conference on Supercomputing},
  pages={289--300},
  year={2023},
  url = {https://doi.org/10.1145/3577193.3593738}
}

@misc{NvidiaDocs,
  author       = {Nvidia},
  title        = {{OptiX Documentation}},
  year = {2026},
  url = {https://raytracing-docs.nvidia.com/optix8/index.html}
}

@inproceedings{gdbod,
  title={{GDBOD: Density-Based Outlier Detection Exploiting Efficient Tree Traversals on the GPU}},
  author={Munugala, Revanth Reddy and Gowanlock, Michael},
  booktitle={2024 IEEE 31st International Conference on High Performance Computing, Data, and Analytics (HiPC)},
  pages={111--121},
  year={2024},
  organization={IEEE},
  url = {https://doi.org/10.1109/HiPC62374.2024.00021}
}

@inproceedings{donnelly2024multi,
  title={{Multi-Space Tree with Incremental Construction for GPU-Accelerated Range Queries}},
  author={Donnelly, Brian and Gowanlock, Michael},
  booktitle={2024 IEEE 31st International Conference on High Performance Computing, Data, and Analytics (HiPC)},
  pages={132--142},
  year={2024},
  organization={IEEE},
  url = {https://doi.org/10.1109/HiPC62374.2024.00024}
}

@misc{FRNN,
  author       = {L. Xue},
  title        = {{Fixed Radius NN Search}},
  year = {2026},
  howpublished = {https://github.com/lxxue/FRNN},
}

@misc{cudamemory,
  author       = {NVIDIA},
  title        = {{CUDA Toolkit Documentation}},
  year = {2026},
  url = {{https://docs.nvidia.com/cuda/cuda-runtime-api/group_\_CUDART\_\_MEMORY.html}}
}

@article{kdTree,
   author  = {Russell A. Brown},
   title   = {{Building a Balanced $k$-d Tree in $O(kn \log n)$ Time}},
   year    = {2015},
   month   = {March},
   day     = {30},
   journal = {Journal of Computer Graphics Techniques (JCGT)},
   volume  = {4},
   number  = {1},
   pages   = {50--68},
   issn    = {2331-7418},
   url = {https://doi.org/10.48550/arXiv.1410.5420}
}

@misc{OWL,
  author = {{Ingo Wald}},
  title  = {{OWL: A Productivity Library for OptiX}},
  year   = {2026},
  note   = {{Available at: \url{https://github.com/NVIDIA/OWL}}}
}

@article{pytorch3d,
    author = {Nikhila Ravi and Jeremy Reizenstein and David Novotny and Taylor Gordon
                  and Wan-Yen Lo and Justin Johnson and Georgia Gkioxari},
    title = {{Accelerating {3D} Deep Learning with PyTorch3D}},
    journal = {arXiv:2007.08501},
    year = {2020},
    url = {https://doi.org/10.48550/arXiv.2007.08501}
}

@InProceedings{PCL,
  author    = {Radu Bogdan Rusu and Steve Cousins},
  title     = {{3D is Here: Point Cloud Library (PCL)}},
  booktitle = {{IEEE International Conference on Robotics and Automation (ICRA)}},
  month     = {May 9-13},
  year      = {2011},
  address   = {Shanghai, China},
  publisher = {IEEE},
  url = {https://doi.org/10.1109/ICRA.2011.5980567}
}

@article{evangelou2021fast,
  title={{Fast Radius Search Exploiting Ray-Tracing Frameworks}},
  author={Evangelou, Iordanis and Papaioannou, Georgios and Vardis, Konstantinos and Vasilakis, Andreas A},
  journal={Journal of Computer Graphics Techniques Vol},
  volume={10},
  number={1},
  pages={25--48},
  year={2021},
  url = {http://jcgt.org/published/0010/01/02/}
}

@article{orair2010distance,
  title={{Distance-Based Outlier Detection: Consolidation and Renewed Bearing}},
  author={Orair, Gustavo H and Teixeira, Carlos HC and Meira Jr, Wagner and Wang, Ye and Parthasarathy, Srinivasan},
  journal={Proceedings of the VLDB Endowment},
  volume={3},
  number={1-2},
  pages={1469--1480},
  year={2010},
  publisher={VLDB Endowment},
  url = {https://doi.org/10.14778/1920841.1921021}
}

@inproceedings{kulkarni2024survey,
  title={{A Survey of Advancements in {DBSCAN} Clustering Algorithms for Big Data}},
  author={Kulkarni, Omkaresh and Burhanpurwala, Adnan},
  booktitle={2024 3rd International Conference on Power Electronics and IoT Applications in Renewable Energy and its Control (PARC)},
  pages={106--111},
  year={2024},
  organization={IEEE},
  url = {https://doi.org/10.1109/PARC59193.2024.10486339}
}

@article{bellman1961adaptive,
  title={{Adaptive Control Processes; A Guided Tour}},
  author={Bellman, R},
  journal={Princeton University Press},
  year={1961},
  url = {https://doi.org/10.2307/3611672}
}

@article{menesesRT,
  title={{Advancing {RT} Core-Accelerated Fixed-Radius Nearest Neighbor Search}},
  author={Meneses, Enzo and Bec, Hugo and Navarro, Crist{\'o}bal A and Crespin, Beno{\^\i}t and Quezada, Felipe A and Hitschfeld, Nancy and Porro, Heinich and Maria, Maxime},
  journal={arXiv preprint arXiv:2601.15633},
  year={2026},
  url = {https://doi.org/10.48550/arXiv.2601.15633}
}

@inproceedings{morton,
  title={{Optimizing Memory Access on {GPUs} using Morton Order Indexing}},
  author={Nocentino, Anthony E and Rhodes, Philip J},
  booktitle={Proceedings of the 48th annual ACM Southeast Conference},
  pages={1--4},
  year={2010},
  url = {https://doi.org/10.1145/1900008.1900035}
}

@article{gowanlock2019accelerating,
  title={{Accelerating the Similarity Self-Join Using the {GPU}}},
  author={Gowanlock, Michael and Karsin, Ben},
  journal={Journal of Parallel and Distributed Computing},
  volume={133},
  pages={107--123},
  year={2019},
  publisher={Elsevier},
  url = {https://doi.org/10.1016/j.jpdc.2019.06.005}
}

@inproceedings{li2015brute,
  title={{Brute-Force k-Nearest Neighbors Search on the GPU}},
  author={Li, Shengren and Amenta, Nina},
  booktitle={International Conference on Similarity Search and Applications},
  pages={259--270},
  year={2015},
  organization={Springer},
  url = {https://doi.org/10.1007/978-3-319-25087-8_25}
}

@inproceedings{mandhare2017comparative,
  title={{A Comparative Study of Cluster Based Outlier Detection, Distance Based Outlier Detection and Density Based Outlier Detection Techniques}},
  author={Mandhare, Harshada C and Idate, SR},
  booktitle={2017 International Conference on Intelligent Computing and Control Systems (ICICCS)},
  pages={931--935},
  year={2017},
  organization={IEEE},
  url = {https://doi.org/10.1109/ICCONS.2017.8250601}
}

@article{schubert2017dbscan,
  title={{{DBSCAN} Revisited, Revisited: Why and How You Should (Still) use {DBSCAN}}},
  author={Schubert, Erich and Sander, J{\"o}rg and Ester, Martin and Kriegel, Hans Peter and Xu, Xiaowei},
  journal={ACM Transactions on Database Systems (TODS)},
  volume={42},
  number={3},
  pages={1--21},
  year={2017},
  publisher={Acm New York, NY, USA},
  url = {https://doi.org/10.1145/3068335}
}

@inproceedings{mathur2024vector,
  title={{Vector Search Algorithms: A Brief Survey}},
  author={Mathur, Srushti and Chhabra, Aayush},
  booktitle={2024 4th International Conference on Ubiquitous Computing and Intelligent Information Systems (ICUIS)},
  pages={365--371},
  year={2024},
  organization={IEEE},
  url = {https://doi.org/10.1109/ICUIS64676.2024.10866377}
}

@misc{nvidia2025blackwell,
  title={{Blackwell GPU Architecture}},
  author={Nvidia},
  journal={NVIDIA Corporation},
  year={2025},
  howpublished={\url{https://images.nvidia.com/aem-dam/Solutions/geforce/blackwell/nvidia-rtx-blackwell-gpu-architecture.pdf}}
}

@incollection{stanzione2020frontera,
  title={{Frontera: The Evolution of Leadership Computing at the National Science Foundation}},
  author={Stanzione, Dan and West, John and Evans, R Todd and Minyard, Tommy and Ghattas, Omar and Panda, Dhabaleswar K},
  booktitle={Practice and Experience in Advanced Research Computing 2020: Catch the Wave},
  pages={106--111},
  year={2020},
  url = {https://doi.org/10.1145/3311790.3396656}
}

@article{wang2025euclidean,
  title={{A Euclidean Distance-Based Novel Algorithm for Binary Feature Selection}},
  author={Wang, Haiyan and Han, Kai and Li, Chuang and Zhao, Jian and Che, Na and Liu, Xiaotong},
  journal={Neural Processing Letters},
  volume={57},
  number={6},
  pages={90},
  year={2025},
  publisher={Springer},
  url = {https://doi.org/10.1007/s11063-025-11797-z}
}

@article{Astronomyreview,
  title={{A Review of Unsupervised Learning in Astronomy}},
  author={Fotopoulou, Sotiria},
  journal={Astronomy and Computing},
  volume={48},
  pages={100851},
  year={2024},
  publisher={Elsevier},
  url = {https://doi.org/10.1016/j.ascom.2024.100851}
}

@incollection{keahey2020lessons,
  title={{Lessons Learned from the Chameleon Testbed}},
  author={Kate Keahey and Jason Anderson and Zhuo Zhen and Pierre Riteau and Paul Ruth and Dan Stanzione and Mert Cevik and Jacob Colleran and Haryadi S. Gunawi and Cody Hammock and Joe Mambretti and Alexander Barnes and Fran\c{c}ois Halbach and Alex Rocha and Joe Stubbs},
  booktitle={Proceedings of the 2020 USENIX Annual Technical Conference (USENIX ATC '20)},
  publisher={USENIX Association},
  month={July},
  year={2020},
  url = {https://dl.acm.org/doi/abs/10.5555/3489146.3489161}
}

@article{gaia2018gaia,
	author = {{Gaia Collaboration} and {Brown, A. G. A.} and {Vallenari, A.} and {Prusti, T.} and {de Bruijne, J. H. J.} and {Babusiaux, C.} and others},
	title = {{Gaia Data Release 2 - Summary of the Contents and Survey Properties}},
	DOI= "10.1051/0004-6361/201833051",
	url= "https://doi.org/10.1051/0004-6361/201833051",
	journal = {Astronomy \& Astrophysics},
	year = 2018,
	volume = 616,
	pages = "A1",
}

@inproceedings{gowanlock2016exploiting,
  title={{Exploiting Variant-Based Parallelism for Data Mining of Space Weather Phenomena}},
  author={Gowanlock, Michael and Blair, David M and Pankratius, Victor},
  booktitle={2016 IEEE International Parallel and Distributed Processing Symposium (IPDPS)},
  pages={760--769},
  year={2016},
  organization={IEEE},
  url = {https://doi.org/10.1109/IPDPS.2016.10}
}

\appendixAD

\section{Overview of Contributions and Artifacts}

\subsection{Paper's Main Contributions}

In this paper, we present the first ray tracing (RT) accelerated distance similarity search algorithm (RT-HiSS) for large high-dimensional datasets. While RT cores are primarily used for graphics applications, several recent works have shown that they can be leveraged for general purpose computation tasks. RT cores are limited to three dimensions due to the infeasibility of representing an object in more than three dimensions in the scene. This is a hardware limitation and requires a software-based solution. In this work, we address this limitation by employing a search-and-refine strategy that enables exploiting the RT cores for indexing using hardware accelerated Bounding Volume Hierarchy (BVH) tree traversals during the search phase and later refining the candidate points using the conventional CUDA cores. We show that this approach exhibits significant performance improvements over the state-of-the-art distance similarity search algorithms. This paper presents the following contributions.

% {\bf MG: General note to Revanth: You're no longer allowed to use the word ``issue'' in your writing :)}
\begin{description}
\item[$C_1$] A RT-accelerated High-dimensional Similarity Search algorithm (RT-HiSS) that addresses the hardware limitations of RT cores.
\item[$C_2$] We optimize BVH tree traversals by identifying and grouping nearby points in the data space together to limit expensive ray-object intersections, particularly for large datasets.
\item[$C_3$] We propose a multi-pass RT approach for our batching scheme that enables near-perfect load balancing across batches with improved caching during the refinement phase.
\item[$C_4$] We address limited GPU memory capacity when the algorithm produces large result sets by employing a compressed result mask strategy, yielding an increased number of queries processed per batch, enabling resource saturation.
\end{description}

\subsection{Computational Artifacts}

The source code of the proposed algorithm (RT-HiSS) can be accessed at $A_1$. Our algorithm requires the OptiX Wrappers Library, which is listed as a Git submodule. The codebase includes a CMakeLists.txt file to compile and run the artifact with a README.md file detailing the installation instructions and various execution options for the algorithm. 

% {\bf NOTE: The zenodo image url doesn't work}
\begin{description}
\item[$A_1$] \url{https://doi.org/10.5281/zenodo.21464379}
\item[$A_2$] \url{https://doi.org/10.5281/zenodo.21504320}
\end{description}

\artexpl{
Provide a table with the relevant computational artifacts,
highlight their relation to the contributions (from above) and
point to the elements in the paper that are reproducible by each artifact, e.g.,
which figures or tables were generated with the artifact.
}

\begin{center}
\begin{tabular}{rll}
\toprule
Artifact ID  &  Contributions &  Related \\
             &  Supported     &  Paper Elements \\
\midrule
$A_1$, $A_2$   &  $C_1$, $C_2$, $C_3$, $C_4$ & Tables III-VI \\
        &        & Figures 6-10\\
 \\
\bottomrule
\end{tabular}
\end{center}

%%%%%%%%%%%%%%%%%%%%%%%%%%%%%%%%%%%%%%%%%%%%%%%%%%%%%%%%
\section{Artifact Identification}
%%%%%%%%%%%%%%%%%%%%%%%%%%%%%%%%%%%%%%%%%%%%%%%%%%%%%%%%

\artexpl{
Provide the following six subsections for each computational artifact $A_i$.
}

\newartifact

\artrel

The relationship between the artifact and contributions is summarized in the table above.

\artexp

% {\bf MG:  Do we need to refer to $C_3$ here?}

Table IV presents the performance of RT-HiSS compared to the five GPU state-of-the-art reference implementations. RT-HiSS should be faster than all reference algorithms across all datasets and search distances except for two evaluation scenarios. These results substantiate $C_1- C_4$. Regarding the performance improvement of individual contributions, grouping nearby points together ($C_2$) and the impact of the compressed result mask strategy ($C_4$) compared to the baseline approaches should be as presented in Figure 7 and Figure 9, respectively.

\arttime

\artexpl{
Estimate the time required to reproduce the artifact, providing separate estimates for the individual steps: Artifact Setup, Artifact Execution, and Artifact Analysis.
}

The artifact setup involves cloning the repository and all its dependencies, and downloading the necessary datasets. This takes roughly 10 minutes. Artifact execution for RT-HISS and all five reference implementations across all datasets and search distances, averaged across three time trials, requires roughly 32 days. Due to the high response time of the reference algorithms in the evaluation, we employ a 60,000 second time out limit for all algorithms. Artifact analysis takes roughly 60 minutes.

\artin

\artinpart{Hardware}

\begin{itemize}
    \item Intel W-2295 CPU with 18 cores.
    \item Quadro RTX 5000 GPU with 16 GiB of global memory.
    \item 256 GiB of main memory.
\end{itemize}

\artinpart{Software}

All experiments are executed on the platform described above with GPU code and host code compiled using CUDA version 13.0, and g++ version 13.3.0, respectively. The CPU code is parallelized using OpenMP 4.5. The OptiX library for RT BVH tree traversals uses version 7.4.0.

\artinpart{Datasets / Inputs}

\artexpl{
Describe the datasets required by the artifact. Indicate whether the datasets can be generated, including instructions, or if they are available for download, providing the corresponding URL.
}

The real world datasets used in the experimental evaluation are listed in Table I, and can be obtained from $A_1$, $A_2$, and at \url{https://rcdata.nau.edu/gowanlock_lab/datasets/scDatasets/}.

\artinpart{Installation and Deployment}

\artexpl{
Detail the requirements for compiling, deploying, and executing the experiments, including necessary compilers and their versions.
}

The software required to reproduce the results is described above. Run the command below to clone the repository with all dependencies. 

\begin{lstlisting}[basicstyle=\small\ttfamily]
git clone --depth 1 --recurse-submodules 
https://github.com/Gowanlock-Lab/rt-hiss.git
\end{lstlisting}

Run the below command to compile and build the artifact.

\begin{lstlisting}[basicstyle=\small\ttfamily]
cmake -S . -B build -DCMAKE_BUILD_TYPE=Release 
-DDIM=<DIM>

cmake --build build -j
\end{lstlisting}
The CMake arguments and other parameters are detailed in the section below.

\artcomp

\artexpl{
Provide an abstract description of the experiment workflow of the artifact. It is important to identify the main tasks (processes) and how they depend on each other.

A workflow may consist of three tasks: $T_1, T_2$, and $T_3$. The task $T_1$ may generate a specific dataset. This dataset is then used as input by a computational task $T_2$, and the output of $T_2$ is processed by another task $T_3$, which produces the final results (e.g., plots, tables, etc.). State the individual tasks $T_i$ and provide their dependencies, e.g., $T_1 \rightarrow T_2 \rightarrow T_3$.

Provide details on the experimental parameters. How and why were parameters set to a specific value (if relevant for the reproduction of an artifact), e.g., size of dataset, number of data points, input sizes, etc. Additionally, include details on statistical parameters, like the number of repetitions.
}

The experiment workflow of the artifact involves two tasks. \noindent{$T_1$}: Loading the dataset into main memory and executing the algorithm. \noindent{$T_2$}: Analyzing the total execution time and the time spent on each component in the algorithm from the \verb|results.txt| file generated at the end of task $T_1$.

All the compile time parameters are passed as CMake arguments. The artifact requires the following compile time parameters.

\begin{itemize}
    \item \Verb[fontsize=\normalsize]{DIM} (default 18): Number of dimensions in the dataset.
    \item \Verb[fontsize=\normalsize]{MAX_KD_LEVELS_OPT} (default: auto): kd-tree height. The default value is $h_{kd} = \lfloor \ln |D| \rfloor$. Here $|D|$ is the total number of points in the dataset.
    \item \Verb[fontsize=\normalsize]{THREADS_TO_COPY_OPT} (default 8): Number of CPU threads for copying results from the device to host.
    \item \Verb[fontsize=\normalsize]{PINNED_MEMORY_SIZE_OPT} (default 16$\times$1024$\times$1024): Allocated pinned buffer size in bytes shared among all CPU threads.
    \item \Verb[fontsize=\normalsize]{NUM_CALCULATIONS_PER_THREAD_OPT} (default 128): Number of point comparisons processed per thread.
    \item \Verb[fontsize=\normalsize]{SAFE_SHARED_MEMORY} (default: auto): Usable shared memory per block in bytes. The default value is set based on the maximum value of shared memory on the GPU.
\end{itemize}

The algorithm requires the following run time parameters passed as command line arguments.

% {\bf NOTE: reorder mode: Maybe state that $(i)$ is the default?}
\begin{itemize}
    \item \Verb[fontsize=\normalsize]{<dataset.txt>}: Path to the dataset file.
    \item \Verb[fontsize=\normalsize]{<epsilon>}: Search distance for distance similarity searches.
    \item \Verb[fontsize=\normalsize]{<reorder mode>} (Optional): Reorders dimensions by variance. Users can select one of (i) highest: highest-variance dimensions first. This is the \textit{default} option. (ii) lowest: lowest-variance dimensions first (iii) random: random dimension order.
\end{itemize}

The datasets and the search distances used in the experiment evaluation are listed in Table I. The compile time argument for dataset dimension (\verb|DIM|) needs to be updated to reflect the dimensionality of the dataset to ensure correctness.

\artout

The artifact analysis is performed using the \verb|results.txt| file. The file contains the configuration options used in the experimental evaluation and allows the user to understand the impact of the proposed optimizations compared to the baseline approaches discussed in the paper. It contains the overall response time, time spent on each algorithmic component (RT time, CUDA time, and CPU/host time). The total number of neighbors across all data points in the datasets is included which can be used to validate algorithm correctness.

%%%%%%%%%%%%%%%%%%%%%%%%%%%%%%%%%%%%%%%%%%%%%%%%%%%%%
%  AE Appendix
%%%%%%%%%%%%%%%%%%%%%%%%%%%%%%%%%%%%%%%%%%%%%%%%%%%%%
\newpage
\appendixAE
% The section numbering is confusing
We provide artifacts $A_1$ and $A_2$ to reproduce the results in the paper. \emph{The most straightforward method for reproducing the results in the paper is to use artifact $A_2$, which is a containerized image.}

{\bf Artifact $A_1$} includes the source files and the \verb|CMakeLists| required to build the artifact, in addition to the scripts to reproduce the results in the paper. 

{\bf Artifact $A_2$} is a containerized image built on the \href{https://chameleoncloud.org/}{Chameleon} platform with all the necessary dependencies, including a compatible CUDA toolkit, OWL, the OptiX SDK, and others, to ease the effort of reproducing the results. 
$A_2$ also contains the source code for \mikealg, which is the most competitive algorithm, as shown in Table~IV in the paper. The source code for the other reference implementations is listed in the paper. We omitted them from $A_2$ due to their high response time. 

% {\bf NOTE: Remember to complete XXX below}

The estimated time to reproduce all the experiments in the paper including setup is roughly 36 hours and 24 minutes using only \wave, \msd, \susy, and \bigcross. We omit the remaining datasets due to their high response times. The estimated time to run each individual experiment is listed in Section~\ref{art:1}. This estimation is based on the Quadro RTX 6000 GPU on the Chameleon platform used in the evaluation and might change depending on the platform. If time is a constraint, interested researchers should focus on comparing the performance of \ouralg with \mikealg. The steps are detailed in Sections~A and B.

\arteval{1}~\label{art:1}
\artin~\label{setup:1}
\artexpl{ % artexpl is the explanatino
Provide instructions for installing and compiling libraries and code.
Offer guidelines on deploying the code to resources.
}

The software for \ouralg can be obtained with the following command.

\begin{lstlisting}[basicstyle=\small\ttfamily]
git clone https://github.com/
Gowanlock-Lab/rt-hiss.git
\end{lstlisting}

\ouralg depends on OWL, a library built on top of OptiX. The OWL library can be initialized as follows.
\begin{lstlisting}[basicstyle=\small\ttfamily]
git submodule update --init --recursive
\end{lstlisting}

The artifact supports numerous CMake options corresponding to the different variants and optimizations discussed in the paper, and it is configured to use the default parameters. To set up the environment, the path to the OptiX SDK must be provided via \verb|-DOptiX_INSTALL_DIR|. The CUDA compiler is detected automatically. Alternatively, the option \verb|-DCMAKE_CUDA_COMPILER| can be used to explicitly set the path to the \verb|nvcc| compiler. The value of \verb|CMAKE_CUDA_ARCHITECTURES| in \verb|CMakeLists| should match the GPU compute capability (for example, 75 for the Turing architecture). The artifact is built with the command below.

\begin{lstlisting}[basicstyle=\small\ttfamily]
cmake -S . -B build \
-DCMAKE_BUILD_TYPE=Release \ 
-DDIM=<num_dimensions> \
-DOptiX_INSTALL_DIR=/path/to/NVIDIA-OptiX-SDK \
-DCMAKE_CUDA_COMPILER=/path/to/nvcc \

cmake --build build -j
\end{lstlisting}

The artifact also allows the following optional compile time arguments to understand the performance of \ouralg using different strategies discussed in the paper.

Below are the options for point comparison variants. These options select where primitive and query points are loaded from (shared vs. global memory). The default loads both from shared memory. 
\Verb[fontsize=\footnotesize]{USE_PRIMITIVE_GLOBAL_QUERY_GLOBAL_NON_BATCHING_OPT} is the baseline where both are loaded from global memory with tiling disabled. Only one point comparison variant can be enabled at a time.

\begin{itemize}
\item \Verb[fontsize=\footnotesize]{USE_PRIMITIVE_SHARED_QUERY_SHARED_BATCHING_OPT}: (default ON): Loads primitive points and query points from shared memory.
\item \Verb[fontsize=\footnotesize]{USE_PRIMITIVE_SHARED_QUERY_GLOBAL_BATCHING_OPT}: Loads primitive points from shared memory and query points from global memory.
\item \Verb[fontsize=\footnotesize]{USE_PRIMITIVE_GLOBAL_QUERY_SHARED_BATCHING_OPT}: Loads primitive points from global memory and query points from shared memory.
\item \Verb[fontsize=\footnotesize]{USE_PRIMITIVE_GLOBAL_QUERY_GLOBAL_BATCHING_OPT}: Loads primitive and query points from global memory.
\end{itemize}

The options below pertain to the result set representation and data copy variants discussed in the paper.

\begin{itemize}
 \item \Verb[fontsize=\footnotesize]{USE_UNCOMPRESSED_MASK_OPT} (default OFF): Use an uncompressed result mask.
 \item  \Verb[fontsize=\footnotesize]{USE_PAGEABLE_MEMORY_OPT} (default OFF): Use pageable memory instead of pinned memory.
 \item \Verb[fontsize=\footnotesize]{USE_CANDIDATE_POINT_COPY_OPT} (default OFF): Use key-value pairs to store the result mask.
\end{itemize}

The options below are available for the ablation study of different workload optimizations.

\begin{itemize}
 \item \Verb[fontsize=\footnotesize]{USE_FIXED_WORK_PER_BLOCK_OPT} (default ON): Enforce a fixed workload per thread block.
 \item \Verb[fontsize=\footnotesize]{USE_SORT_BY_WORKLOAD_OPT} (default ON): Sort workloads from lowest to highest to reduce tail effects.

\end{itemize}

The following example shows how to build the artifact with several of the parameter configurations discussed above. Specifically, it builds the artifact with the default point comparison variant and data copy variant, together with default values for the number of CPU threads, pinned memory, and shared memory. The software requires a GPU with ray tracing cores and we describe using the Quadro RTX 6000 as a target GPU as this is the only ray tracing capable GPU on the Chameleon platform at the time of writing the paper. 

\begin{lstlisting}[basicstyle=\small\ttfamily]
cmake -S . -B build \
  -DCMAKE_BUILD_TYPE=Release \ 
  -DOptiX_INSTALL_DIR=/path/to/NVIDIA-OptiX-SDK \
  -DCMAKE_CUDA_COMPILER=/path/to/nvcc \
  -DDIM=2 \
  -DUSE_PRIMITIVE_SHARED_QUERY_SHARED_BATCHING
  _OPT=ON \
  -DMAX_KD_LEVELS_OPT=12 \
  -DTHREADS_TO_COPY_OPT=8 \
  -DPINNED_MEMORY_SIZE_OPT=16777216 \
  -DSAFE_SHARED_MEMORY=48768 \
  -DUSE_UNCOMPRESSED_MASK_OPT=OFF \
  -DUSE_PAGEABLE_MEMORY_OPT=OFF \
  -DUSE_USE_CANDIDATE_POINT_COPY_OPT=OFF

cmake --build build -j
\end{lstlisting}

The OptiX SDK is available at \url{https://developer.nvidia.com/designworks/optix/downloads/legacy}, and the NVIDIA CUDA toolkit is available at \url{https://developer.nvidia.com/cuda/toolkit}.
The datasets must be in CSV format, containing numerical values and omit a header. Each row is a feature vector in the dataset with a comma separating each dimension. A sample dataset is available  in \verb|sampleDataset| directory. The number of dimensions in a dataset must match the \verb|DIM| parameter specified at build time. The datasets used in the paper are stored with \href{https://git-lfs.com/}{Git LFS} (Large File Storage), which is therefore required to download them. Git LFS can be installed as follows.

\begin{lstlisting}[basicstyle=\small\ttfamily]
sudo apt install git-lfs
\end{lstlisting}

After installation, Git LFS needs to be set up once, which is done by executing the command below.

\begin{lstlisting}[basicstyle=\small\ttfamily]
git lfs install 
\end{lstlisting}

The datasets, provided as a gzipped tar file, are then downloaded with the following command.

\begin{lstlisting}[basicstyle=\small\ttfamily]
git lfs pull 
\end{lstlisting}

The tar file (\verb|scDatasets.tar.gz|) can be unpacked as follows.

\begin{lstlisting}[basicstyle=\small\ttfamily]
tar -xvzf scDatasets.tar.gz
\end{lstlisting}

Alternatively, the datasets used in the paper are also available at \url{https://rcdata.nau.edu/gowanlock_lab/datasets/scDatasets}.

\artcomp~\label{analysis-1}

\artexpl{
Describe the experiment workflow.
If encapsulated within a workflow description or equivalent (such as a makefile or script), clearly outline the primary tasks and their interdependencies. Detail the main steps in the workflow. Merely instructing to “Run script.sh” is inadequate.
}

The artifact requires three runtime arguments: (i) the dataset path, (ii) the search distance (epsilon), and (iii) the reorder mode (optional). \ouralg can be run using the command format shown below.

\begin{lstlisting}[basicstyle=\small\ttfamily]
./build/RT-HiSS <dataset.txt> \
    <epsilon> [reorder_mode]
\end{lstlisting}

A sample example is given below.

\begin{lstlisting}[basicstyle=\small\ttfamily]
./build/RT-HiSS sampleDataset/iono_57_1000.txt \
    0.01
\end{lstlisting}

\artout

We provide bash scripts in the \verb|experiments| directory to simplify reproducing the results and generating the plots used in the paper. Specifically, we include scripts to reproduce Figures 7-10 and Table~IV. We omit Figure~6 because parameter tuning is a time-consuming step. However, it can be reproduced by explicitly setting the \verb|-DMAX_KD_LEVELS_OPT| parameter in the build step. By default, all of the scripts use only the \wave, \msd, \susy, and \bigcross datasets and omit the others, as the remaining datasets require considerably more time and would severely increase the total time needed to reproduce the results. Where applicable, all of the scripts generate plots to validate and compare the results obtained on the evaluation platform with those reported in the paper. The total time to reproduce all experiments is 36 hours and 24 minutes. 

\subsubsection{\ouralg vs. \mikealg (Table~IV)}~\label{subsubsec:sota_art1} 

% {\bf NOTE: remember to complete xx below.}

Table~IV compares the performance of \ouralg with that of state-of-the-art (SOTA) reference algorithms. The script \verb|run_default.sh| builds the artifact for each dataset using the default parameters and executes \ouralg with the search distances corresponding to the small, medium, and large selectivities listed in Table~I of the paper. For all datasets, the build parameters for each run configuration and the total time, including the time for each component, are appended to the \verb|results_default.txt| file. The expected time for reproducing this experiment is roughly 1 hour and 5 minutes.

The source code for \mikealg  includes a \verb|Makefile| for compilation, along with instructions for running the algorithm and can be obtained with the following command.
\begin{lstlisting}[basicstyle=\small\ttfamily]
git clone https://github.com/mgowanlock/
gpu_self_join.git
\end{lstlisting}

% Add textbf?
\noindent{Result expectation:} \ouralg should be considerably faster than \mikealg across most datasets. It may, however, be slower than \mikealg on the \susy dataset with the search distance with smallest selectivity.

% , since the end-to-end time there is roughly 1 second and the costs of OptiX setup and other overheads are fixed regardless of the dataset.

% {\bf NOTE: remember to complete xx below.}
\subsubsection{Impact of grouping points together (Figure~7)} 

Figure~7 compares the performance of \ouralg, in which a primitive is defined by the combined $\epsilon$-neighborhood of nearby points, against a variant of \ouralg in which the $\epsilon$-neighborhood of each point is represented as an individual primitive. The script \verb|run_one_point.sh| builds two variants of the artifact by adjusting the \verb|-DMAX_KD_LEVELS_OPT| parameter. The total time required for each dataset across all search distances is appended to \verb|results_one_point.txt|, and a plot comparing both variants is generated at \verb|plot/one_point.pdf|. The expected time for this experiment is 34 hours and 20 minutes.

\noindent{(i) Nearby points are grouped together:}  Here, \verb|-DMAX_KD_LEVELS_OPT = |$\lfloor \ln |D| \rfloor$ $(h_{kd})$. As discussed in the paper, the height of the kd-tree affects the number of primitives. Specifically, the OptiX scene contains $2^{h_{kd}-1}$ primitives, with roughly ${|D|}/{2^{h_{kd} -1}}$ nearby points grouped together in each primitive.

The following is a sample build of \ouralg for \wave when nearby points are grouped together using the heuristic above.

\begin{lstlisting}[basicstyle=\small\ttfamily]
cmake -S . -B build \
    -DDIM = 49 \
    -DMAX_KD_LEVELS_OPT=12 \

cmake --build build -j
\end{lstlisting}

\noindent{(ii) $\epsilon$-neighborhood of each point as an individual primitive: } This is achieved by adjusting the value of $h_{kd}$ such that ${|D|}/{2^{h_{kd}-1}} = 1$.

The following is a sample build of \ouralg for \wave when each point is represented as a primitive.

\begin{lstlisting}[basicstyle=\small\ttfamily]
cmake -S . -B build \
    -DDIM = 49 \
    -DMAX_KD_LEVELS_OPT=19 \
cmake --build build -j
\end{lstlisting}

\noindent{Result expectation:} \ouralg should exhibit a significant performance improvement over the variant that represents the $\epsilon$-neighborhood of each point as a primitive, across all datasets and search distances. The generated plot should resemble Figure~7 in the paper.

\subsubsection{\ouralg vs. Point Comparison Variants (Figure~8)} 

Figure~8 illustrates the performance of \ouralg relative to different point comparison variants. The script \verb|run_batching.sh| builds the artifact and generates a plot at \verb|plots/batching.pdf| comparing the end-to-end performance of different point comparison variants for each dataset and search distance detailed in the paper. The expected time for reproducing this experiment is 55 minutes.

\noindent{(i)} \shqp (\ouralg): This is the default configuration of \ouralg, in which both primitive and query points are staged and read from shared memory.

\begin{lstlisting}[basicstyle=\small\ttfamily]
cmake -S . -B build \
    -DDIM = 49 \
    -DUSE_PRIMITIVE_SHARED_QUERY_SHARED_BATCHING
    _OPT=ON \ 

cmake --build build -j
\end{lstlisting}

\noindent{(ii)} \shp: This is a variant of \ouralg in which only primitive points are read from shared memory.

\begin{lstlisting}[basicstyle=\small\ttfamily]
cmake -S . -B build \
    -DDIM = 49 \
    -DUSE_PRIMITIVE_SHARED_QUERY_GLOBAL_BATCHING
    _OPT=ON \ 

cmake --build build -j
\end{lstlisting}

\noindent{(iii)} \shq: This is a variant of \ouralg where only query points are read from shared memory.

\begin{lstlisting}[basicstyle=\small\ttfamily]
cmake -S . -B build \
    -DDIM = 49 \
    -DUSE_PRIMITIVE_GLOBAL_QUERY_SHARED_BATCHING
    _OPT=ON \ 

cmake --build build -j
\end{lstlisting}

\noindent{(iv)} \shno: This is the baseline variant, in which both query and primitive points are read from global memory without being staged in shared memory.

\begin{lstlisting}[basicstyle=\small\ttfamily]
cmake -S . -B build \
    -DDIM = 49 \
    -DUSE_PRIMITIVE_GLOBAL_QUERY_GLOBAL_BATCHING
    _OPT=ON \ 

cmake --build build -j
\end{lstlisting}

\noindent{Result expectation:} \shqp (\textit{default}) should exhibit significant performance improvements over the baseline variant (\shno), and it should be faster than the other variants in most scenarios. In a few cases, \shp may perform better than \shqp. The generated plot should resemble Figure~8 in the paper.

\subsubsection{\ouralg vs. Result Representation Variants (Figure~9)} 

Figure~9 compares the performance of \ouralg across the different result representation variants discussed in the paper. The script \verb|run_data_copy.sh| builds the artifact with the following variants and generates a plot at \verb|plots/data_copy.pdf| that compares the performance of different result representation variants. The expected time for reproducing this experiment is 45 minutes.

\noindent{(i) \textsc{Compressed Result Mask} (\ouralg): } This is the default approach, in which the result set is represented using a mask that is compressed to mitigate the cost of transferring a large result set from device to host.

\begin{lstlisting}[basicstyle=\small\ttfamily]
cmake -S . -B build \
    -DDIM = 49 \
    -DUSE_UNCOMPRESSED_MASK_OPT=OFF \
    -DUSE_USE_CANDIDATE_POINT_COPY_OPT=OFF \

cmake --build build -j
\end{lstlisting}

\noindent{(ii) \textsc{Uncompressed Result Mask}: } This is a variant of \ouralg in which compression is not applied to the result mask, and the uncompressed mask is transferred from device to host.

\begin{lstlisting}[basicstyle=\small\ttfamily]
cmake -S . -B build \
    -DDIM = 49 \
    -DUSE_UNCOMPRESSED_MASK_OPT=ON \
    -DUSE_USE_CANDIDATE_POINT_COPY_OPT=OFF \

cmake --build build -j
\end{lstlisting}

\noindent{(iii) \textsc{Key-Value Copy}: } This is a variant of \ouralg in which the result set is stored using traditional key-value pairs instead of the compressed result mask.

\begin{lstlisting}[basicstyle=\small\ttfamily]
cmake -S . -B build \
    -DDIM = 49 \
    -DUSE_USE_CANDIDATE_POINT_COPY_OPT=ON \

cmake --build build -j
\end{lstlisting}

\noindent{Result expectation:} \textsc{Compressed Result Mask} (\ouralg) should be faster than the other variants, and the generated plot should resemble Figure~9 in the paper.

\subsubsection{Time Distribution Analysis (Figure~10)} 

Figure~10 is used to examine the time spent by each component of \ouralg using the default parameters. Specifically, \verb|run_time_distribution.sh| investigates the time spent on the first intersection test, the total RT core time, the total CUDA core time, the compression time, the kd-tree construction time, and the CPU time for the remaining tasks. A plot illustrating the average time spent by each component for each dataset, across all search distances is generated at \verb|plots/time_distribution.pdf|. The expected time for this experiment is 12 minutes.

\noindent{Result expectation:} The plot should resemble Figure~10 in the paper. The majority of the time should be spent on CUDA core refinement and kd-tree construction. Only a minimal amount of time should be spent on identifying candidate neighbor points using RT cores, compressing the result mask, and copying the result set from device to host.

\artexpl{
\begin{itemize}
    \item Provide a description of the expected results and a methodology for evaluating these results.
    \item Explain how the expected results from the experiment workflow correlate with the contributions stated in the article.
    \item For example, if the article presents results in a figure, the artifact evaluation should also produce a similar figure, depicting the same generalizable outcome. Authors must focus on these aspects to reduce the time required for others to understand and verify an artifact.
\end{itemize}
}

% Investigate extra space
\arteval{2}~\label{art:2}

\artin % Artifact setup

Artifact $A_2$ contains a self-contained image with the required datasets and all the dependencies, including the OWL SDK, OptiX SDK, NVIDIA CUDA toolkit, and every other package needed to build and run the artifact. Artifact $A_2$ is built on the Chameleon platform.

On the \href{https://chameleoncloud.org/}{Chameleon} platform, reserve a node with a ray-tracing-capable GPU (Quadro RTX 6000) on the CHI@UC site. At the time of writing, the image is not available at other Chameleon sites, such as CHI@TACC and CHI@NCAR, since these sites do not provide the compatible GPU with RT cores required to reproduce the artifact. Create an instance using the image \verb|sc_ae_rt-hiss_22072026| from the \href{https://chameleoncloud.org/appliances/134/}{Chameleon appliance catalog} or simply use the launch option from the \href{https://trovi.chameleoncloud.org/dashboard/artifacts/83418cdd-aac5-4065-b6a1-58fb417aa922}{Trovi} repository. Further instructions on creating an instance from the image are available on the support page \href{https://chameleoncloud.readthedocs.io/en/latest/technical/baremetal/launching_gui.html}{here}. It takes roughly 12 minutes to create an instance from the image.

\artcomp~\label{sec:sota_art2}

The steps outlined in Section~A apply here as well for reproducing Figures 7-10. To reproduce Table~IV, use \verb|experiments/run_default.sh| to run \ouralg with the default parameters. The image also includes the source code and the relevant dependencies for the most competitive algorithm, \mikealg. Run \mikealg with the default parameters on all datasets using \verb|gds-join/run_all_gds-join.sh|. The total time for each dataset and search distance is stored in \verb|gpu_stats.txt|.

\artout % Artifact Analysis

The image includes a \verb|compare_times.sh| bash script that compares the performance of \ouralg relative to \mikealg across all datasets. This script should be executed after executing \verb|experiments/run_default.sh| for \ouralg and \verb|gds-join/run_all_gds-join.sh| for \mikealg. The results will be similar to Table~IV in the paper. To reproduce the results for the remaining experiments, the steps outlined in Section~A apply here as well.

\end{document}